\documentclass[12pt]{article}
\usepackage[cmtip,arrow]{xy}\usepackage{pb-diagram,pb-xy}%

\input xy
\xyoption{all}
\input xy
\xyoption{all}
\usepackage{heck}
\usepackage{cite}
\usepackage{graphicx}
\usepackage{makeidx}
\usepackage{multicol}
\usepackage{geometry}
\usepackage{amsfonts}
\usepackage{mathrsfs}
\usepackage{amssymb}
\usepackage{amsmath}%

\usepackage{lscape}

\providecommand{\U}[1]{\protect\rule{.1in}{.1in}}
\numberwithin{equation}{section}

\newcommand{\bea}{\begin{eqnarray}}
\newcommand{\eea}{\end{eqnarray}}
\newcommand{\be}{\begin{equation}}
\newcommand{\ee}{\end{equation}}

\newcommand{\bem}{\begin{pmatrix}}
\newcommand{\eem}{\end{pmatrix}}

\def\a{\alpha}

\def\d{\delta}

\def\s{\sigma}                                   

\def\U{\Upsilon}

\def\cn{{\cal N}}

\def \N {{\mathcal N}}

\xyoption{arc}

\begin{document}

\bibliographystyle{utphys}

\date{October, 2011}


\institution{SISSA}{\centerline{${}$SISSA, Scuola Internazionale Superiore di Studi Avanzati,
I-34100 Trieste, ITALY}}

\title{More Arnold's $\cn=2$ superconformal gauge theories}
%

\authors{Michele Del Zotto\worksat{\SISSA}\footnote{e-mail: {\tt eledelz@gmail.com}}}%


\abstract{We study the $\cn = 2$ gauge theories obtained by engineering the Type IIB superstring on the quasi--homogeneous elements of Arnold's list of bimodal singularities. All these theories have \emph{finite} BPS chambers and we describe, along the lines of \texttt{arXiv:1107.5747}, the algebraically obvious ones.

\smallskip

Our results leads to the prediction of 11 new \emph{periodic} $Y$--systems, providing additional evidence for the correspondence in between thermodinamical Bethe ansatz periodic $Y$--systems and $\cn = 2$ superconformal theories with a finite BPS chamber whose chiral primaries have dimensions of the form $\mathbb{N}/\ell$.}


\maketitle

\tableofcontents

\section{Introduction}
The study of the wall--crossing phenomenon in 4--dimensional $\cn = 2$ supersymmetric quantum field theories\cite{SW1,SW2,KS,DG,DGS,GMN:2008,GMN:2009} is unravelling deep relations \cite{CV09,CVN,CDZ} in between BPS spectroscopy and quantum cluster algebras \cite{Cluster}  for the class of $\cn = 2$ theories that admit BPS quiver \cite{CV11}. In particular, the fundamental wall--crossing invariant, the $4d$ quantum monodromy $\mathbb{M}(q)$, admits various distinct factorizations into basic quantum mutations each corresponding to a \emph{formal} stable BPS spectrum --- see \cite{CVN,CDZ,Keller:2011} and references therein. BPS quiver theories divides into two classes: complete theories\cite{CV11,CVal:complete}, for which all possible BPS chambers are physically realized, and non--complete ones, for which the physical submanifold is a proper subset of the space of stability conditions --- a fact that leads to the so called \emph{quantum Schottky problem}.

\medskip

In ref.\cite{CDZ}, an extension of the CNV strategy to find such factorizations was obtained to handle models that are more general than square tensor ones \cite{CVN,Keller:DynkinPairs} and it was used to study the non--square tensor models out of the Arnold's exceptional unimodal singularities list \cite{AGV,EMS,Ebeling1}. Unimodal $\cn = 2$ superconformal theories are the simplest non--complete ones, the physical submanifold having codimension one. The purpose of this note is to study, along the lines of \cite{CDZ}, the further layer of non--completeness, the models obtained by type IIB engineering on quasi--homogenous bimodal singularities.

\medskip

The structure of this paper is the following: In section 2 we discuss the quasi--homogeneous elements of Arnold's bimodal singularities list. In section 3 we discuss (mostly) the information about the models coming from the $2d$ theory: the issue of non--completeness and its relation with the modality of a singularity \cite{AGV,EMS,Ebeling1,Ebeling2,Gabrielov2,Gabrielov3,GabrielovKushnirenko} are discussed in section 3.1; the r\^ole of the $2d$ wall--crossing group \cite{CV92,CV92bis,WClectures,HIV} is presented in \S 3.2, manifesting itself for the even elements of the $Q$ series; in \S 3.3 we discuss, by the RG argument of \cite{CDZ}, the issue of the physical realizableness of the `strongly coupled' BPS chambers we have obtained; in \S 3.4 we compute the number of flavor charges of the bimodal superconformal theories. In section 4, we consider the BPS spectroscopy at `strong coupling'. We refer to \cite{CDZ} for the explanation of the (extended) CNV strategy --- see in particular \S 4,5 --- and we limit ourselves to stating our results, referring to appendix A for the details of  computations. The delicate interplay in between cluster algebras theory and the thermodynamical Bethe ansatz --- see \cite{Keller:DynkinPairs,Keller:Triang,FZ:Ys} and references therein --- leads, as a side result of this work, to the prediction of 11 new periodic $Y$--systems, we briefly discuss this fact in \S 5. All Coxeter--Dynkin diagrams we have used here were obtained by A. M. Gabrielov in \cite{Gabrielov, Gabrielov2} --- see also \cite{Ebeling2}.

\section{Bimodal singularities}

\begin{table}
\caption{Exceptional bimodal singularities that are not direct sum of simple ones.}\label{ArnB1}
\begin{center}
\begin{tabular}{|c|c|c|}\hline
name & $W(x,y,z)$ & Coxeter--Dynkin diagram\\
\hline
$\begin{matrix}\\ E_{19}\\ \phantom{a}\end{matrix}$ & $\begin{matrix}\\ x^3+xy^7+z^2\\ \phantom{a}\end{matrix}$ & {\scriptsize $\begin{gathered}\xymatrix{
\bullet \ar@{-}[r]\ar@{-}[d]\ar@{..}[dr]&\bullet \ar@{-}[r]\ar@{-}[d]\ar@{..}[dr]&\bullet \ar@{-}[r]\ar@{-}[d]\ar@{..}[dr]&\bullet \ar@{-}[r]\ar@{-}[d]\ar@{..}[dr]&\bullet \ar@{-}[r]\ar@{-}[d]\ar@{..}[dr]&\bullet \ar@{-}[d]\ar@{..}[dr]\ar@{-}[r]&\bullet \ar@{-}[d]\ar@{..}[dr]\ar@{-}[r]&\bullet \ar@{-}[d]\ar@{..}[dr]\ar@{-}[r]&\bullet\ar@{-}[d]\ar@{..}[dr]\\
\bullet \ar@{-}[r]&\bullet \ar@{-}[r]&\bullet \ar@{-}[r]&\bullet \ar@{-}[r]&\bullet \ar@{-}[r]&\bullet\ar@{-}[r]&\bullet\ar@{-}[r]&\bullet\ar@{-}[r]&\bullet\ar@{-}[r]&\bullet }\end{gathered}$}\\\hline
$\begin{matrix}\\ Z_{17}\\ \phantom{a}\end{matrix}$ & $\begin{matrix}\\ x^3y+y^8+z^2\\ \phantom{a}\end{matrix}$ & {\scriptsize $\begin{gathered}\xymatrix{
\bullet \ar@{-}[r]\ar@{-}[d]&\bullet \ar@{-}[r]\ar@{-}[d]\ar@{..}[dl]&\bullet \ar@{-}[d]\ar@{..}[dl] & &\\
\bullet \ar@{-}[r]\ar@{-}[d]\ar@{..}[dr]&\bullet \ar@{-}[r]\ar@{-}[d]\ar@{..}[dr]&\bullet \ar@{-}[r]\ar@{-}[d]\ar@{..}[dr]&\bullet \ar@{-}[r]\ar@{-}[d]\ar@{..}[dr]&\bullet \ar@{-}[r]\ar@{-}[d]\ar@{..}[dr]&\bullet \ar@{-}[d] \ar@{..}[dr]\ar@{-}[r]&\bullet \ar@{-}[d]\\
\bullet \ar@{-}[r]&\bullet \ar@{-}[r]&\bullet \ar@{-}[r]&\bullet \ar@{-}[r]&\bullet \ar@{-}[r]&\bullet \ar@{-}[r]&\bullet\\
}\end{gathered}$}\\\hline
$\begin{matrix}\\ Z_{18}\\ \phantom{a}\end{matrix}$ & $\begin{matrix}\\ x^3y+xy^6+z^2\\ \phantom{a}\end{matrix}$ & {\scriptsize $\begin{gathered}\xymatrix{
\bullet \ar@{-}[r]\ar@{-}[d]&\bullet \ar@{-}[r]\ar@{-}[d]\ar@{..}[dl]&\bullet \ar@{-}[d]\ar@{..}[dl] & &\\
\bullet \ar@{-}[r]\ar@{-}[d]\ar@{..}[dr]&\bullet \ar@{-}[r]\ar@{-}[d]\ar@{..}[dr]&\bullet \ar@{-}[r]\ar@{-}[d]\ar@{..}[dr]&\bullet \ar@{-}[r]\ar@{-}[d]\ar@{..}[dr]&\bullet \ar@{-}[r]\ar@{-}[d]\ar@{..}[dr]&\bullet \ar@{-}[d] \ar@{..}[dr]\ar@{-}[r]&\bullet \ar@{-}[d]\ar@{..}[dr]\\
\bullet \ar@{-}[r]&\bullet \ar@{-}[r]&\bullet \ar@{-}[r]&\bullet \ar@{-}[r]&\bullet \ar@{-}[r]&\bullet \ar@{-}[r]&\bullet\ar@{-}[r]&\bullet\\
}\end{gathered}$}\\\hline
$\begin{matrix}\\ Z_{19}\\ \phantom{a}\end{matrix}$ & $\begin{matrix}\\ x^3y+y^9+z^2\\ \phantom{a}\end{matrix}$ &  {\scriptsize $\begin{gathered}\xymatrix{
\bullet \ar@{-}[r]\ar@{-}[d]&\bullet \ar@{-}[r]\ar@{-}[d]\ar@{..}[dl]&\bullet \ar@{-}[d]\ar@{..}[dl] & &\\
\bullet \ar@{-}[r]\ar@{-}[d]\ar@{..}[dr]&\bullet \ar@{-}[r]\ar@{-}[d]\ar@{..}[dr]&\bullet \ar@{-}[r]\ar@{-}[d]\ar@{..}[dr]&\bullet \ar@{-}[r]\ar@{-}[d]\ar@{..}[dr]&\bullet \ar@{-}[r]\ar@{-}[d]\ar@{..}[dr]&\bullet \ar@{-}[d] \ar@{..}[dr]\ar@{-}[r]&\bullet \ar@{-}[d] \ar@{..}[dr]\ar@{-}[r]&\bullet \ar@{-}[d]\\
\bullet \ar@{-}[r]&\bullet \ar@{-}[r]&\bullet \ar@{-}[r]&\bullet \ar@{-}[r]&\bullet \ar@{-}[r]&\bullet \ar@{-}[r]&\bullet \ar@{-}[r]&\bullet\\
}\end{gathered}$}\\\hline
$\begin{matrix}\\ W_{17}\\ \phantom{a}\end{matrix}$ & $\begin{matrix}\\ x^4+x y^5+z^2\\ \phantom{a}\end{matrix}$ &  {\scriptsize $\begin{gathered}\xymatrix{\bullet \ar@{-}[r]\ar@{-}[d]&\bullet \ar@{-}[r]\ar@{-}[d]\ar@{..}[dl]&\bullet \ar@{-}[d]\ar@{..}[dl]\ar@{-}[r]&\bullet \ar@{-}[d]\ar@{..}[dl]\ar@{-}[r]& \bullet \ar@{-}[d]\ar@{..}[dl]\\
\bullet \ar@{-}[r]\ar@{-}[d]\ar@{..}[dr]&\bullet \ar@{-}[r]\ar@{-}[d]\ar@{..}[dr]&\bullet \ar@{-}[r]\ar@{-}[d]\ar@{..}[dr]&\bullet \ar@{-}[r]\ar@{-}[d]\ar@{..}[dr]&\bullet \ar@{-}[r]\ar@{-}[d]\ar@{..}[dr]&\bullet \ar@{-}[d]\\
\bullet \ar@{-}[r]&\bullet \ar@{-}[r]&\bullet \ar@{-}[r]&\bullet \ar@{-}[r]&\bullet \ar@{-}[r]&\bullet\\
}\end{gathered}$}\\\hline
\end{tabular}
\end{center}
\end{table}

\begin{table}
\begin{center}
\begin{tabular}{|c|c|c|}\hline
$\begin{matrix}\\ Q_{16}\\ \phantom{a}\end{matrix}$ & $\begin{matrix}\\ x^3+yz^2+y^7\\ \phantom{a}\end{matrix}$ &{\scriptsize$\begin{gathered}\xymatrix{\bullet\ar@{-}[r]\ar@{-}@/_/[dd]&\bullet \ar@{..}[ddl]\ar@{-}@/_/[dd]&&\\
\bullet \ar@{-}[r]\ar@{-}[d]&\bullet \ar@{-}[d]\ar@{..}[dl]& &\\
\bullet \ar@{-}[r]\ar@{-}[d]\ar@{..}[dr]&\bullet \ar@{-}[r]\ar@{-}[d]\ar@{..}[dr]&\bullet \ar@{-}[r]\ar@{-}[d]\ar@{..}[dr]&\bullet \ar@{-}[r]\ar@{-}[d]\ar@{..}[dr]&\bullet \ar@{-}[r]\ar@{-}[d]\ar@{..}[dr]&\bullet\ar@{-}[d] \\
\bullet \ar@{-}[r]&\bullet \ar@{-}[r]&\bullet \ar@{-}[r]&\bullet \ar@{-}[r]&\bullet \ar@{-}[r]&\bullet\\
}\end{gathered}$}\\\hline
$\begin{matrix}\\ Q_{17}\\ \phantom{a}\end{matrix}$ & $\begin{matrix}\\ x^3+yz^2+xy^5\\ \phantom{a}\end{matrix}$ &{\scriptsize$\begin{gathered}\xymatrix{\bullet\ar@{-}[r]\ar@{-}@/_/[dd]&\bullet \ar@{..}[ddl]\ar@{-}@/_/[dd]&&\\
\bullet \ar@{-}[r]\ar@{-}[d]&\bullet \ar@{-}[d]\ar@{..}[dl]& &\\
\bullet \ar@{-}[r]\ar@{-}[d]\ar@{..}[dr]&\bullet \ar@{-}[r]\ar@{-}[d]\ar@{..}[dr]&\bullet \ar@{-}[r]\ar@{-}[d]\ar@{..}[dr]&\bullet \ar@{-}[r]\ar@{-}[d]\ar@{..}[dr]&\bullet \ar@{-}[r]\ar@{-}[d]\ar@{..}[dr]&\bullet\ar@{-}[d]\ar@{..}[dr] \\
\bullet \ar@{-}[r]&\bullet \ar@{-}[r]&\bullet \ar@{-}[r]&\bullet \ar@{-}[r]&\bullet \ar@{-}[r]&\bullet \ar@{-}[r]&\bullet\\
}\end{gathered}$}\\\hline
$\begin{matrix}\\ Q_{18}\\ \phantom{a}\end{matrix}$ & $\begin{matrix}\\ x^3+yz^2+y^8\\ \phantom{a}\end{matrix}$ &{\scriptsize$\begin{gathered}\xymatrix{\bullet\ar@{-}[r]\ar@{-}@/_/[dd]&\bullet \ar@{..}[ddl]\ar@{-}@/_/[dd]&&\\
\bullet \ar@{-}[r]\ar@{-}[d]&\bullet \ar@{-}[d]\ar@{..}[dl]& &\\
\bullet \ar@{-}[r]\ar@{-}[d]\ar@{..}[dr]&\bullet \ar@{-}[r]\ar@{-}[d]\ar@{..}[dr]&\bullet \ar@{-}[r]\ar@{-}[d]\ar@{..}[dr]&\bullet \ar@{-}[r]\ar@{-}[d]\ar@{..}[dr]&\bullet \ar@{-}[r]\ar@{-}[d]\ar@{..}[dr]&\bullet \ar@{-}[r]\ar@{-}[d]\ar@{..}[dr]&\bullet\ar@{-}[d] \\
\bullet \ar@{-}[r]&\bullet \ar@{-}[r]&\bullet \ar@{-}[r]&\bullet \ar@{-}[r]&\bullet \ar@{-}[r]&\bullet \ar@{-}[r]&\bullet\\
}\end{gathered}$}\\\hline
$\begin{matrix}\\ S_{16}\\ \phantom{a}\end{matrix}$ & $\begin{matrix}\\ x^2z+yz^2+xy^4\\ \phantom{a}\end{matrix}$ & {\scriptsize$\begin{gathered}\xymatrix{\bullet\ar@{-}[r]\ar@{-}@/_/[dd]&\bullet \ar@{..}[ddl]\ar@{-}@/_/[dd]&\\
\bullet \ar@{-}[r]\ar@{-}[d]&\bullet \ar@{-}[d]\ar@{..}[dl]\ar@{-}[r]&\bullet \ar@{-}[d]\ar@{..}[dl]\ar@{-}[r]&\bullet \ar@{-}[d]\ar@{..}[dl]&\\
\bullet \ar@{-}[r]\ar@{-}[d]\ar@{..}[dr]&\bullet \ar@{-}[r]\ar@{-}[d]\ar@{..}[dr]&\bullet \ar@{-}[r]\ar@{-}[d]\ar@{..}[dr]&\bullet \ar@{-}[r]\ar@{-}[d]\ar@{..}[dr]&\bullet \ar@{-}[d] \\
\bullet \ar@{-}[r]&\bullet \ar@{-}[r]&\bullet \ar@{-}[r]&\bullet \ar@{-}[r]&\bullet\\
}\end{gathered}$}\\\hline
$\begin{matrix}\\ S_{17}\\ \phantom{a}\end{matrix}$ & $\begin{matrix}\\ x^2z+yz^2+y^6\\ \phantom{a}\end{matrix}$ & {\scriptsize$\begin{gathered}\xymatrix{\bullet\ar@{-}[r]\ar@{-}@/_/[dd]&\bullet \ar@{..}[ddl]\ar@{-}@/_/[dd]&\\
\bullet \ar@{-}[r]\ar@{-}[d]&\bullet \ar@{-}[d]\ar@{..}[dl]\ar@{-}[r]&\bullet \ar@{-}[d]\ar@{..}[dl]\ar@{-}[r]&\bullet \ar@{-}[d]\ar@{..}[dl]\ar@{-}[r]&\bullet \ar@{-}[d]\ar@{..}[dl]\\
\bullet \ar@{-}[r]\ar@{-}[d]\ar@{..}[dr]&\bullet \ar@{-}[r]\ar@{-}[d]\ar@{..}[dr]&\bullet \ar@{-}[r]\ar@{-}[d]\ar@{..}[dr]&\bullet \ar@{-}[r]\ar@{-}[d]\ar@{..}[dr]&\bullet \ar@{-}[d] \\
\bullet \ar@{-}[r]&\bullet \ar@{-}[r]&\bullet \ar@{-}[r]&\bullet \ar@{-}[r]&\bullet\\
}\end{gathered}$}\\\hline
\end{tabular}
\end{center}
\end{table}

\begin{table}
\caption{Quasi--homogeneous elements of the 8 infinite series of bimodal singularities that are not direct sum of simple ones. We indicate the corresponding Milnor numbers in parenthesis.}\label{ArnB2}
\begin{center}
\begin{tabular}{|c|c|c|}\hline
name & $W(x,y,z)$ & Coxeter--Dynkin diagram\\
\hline
$\begin{matrix}\\ Z_{1,0}  \\ \phantom{a}\end{matrix}$ {\footnotesize (15)} & $\begin{matrix}\\ x^3 y+y^7+z^2 \\ \phantom{a}\end{matrix}$ & {\scriptsize $\begin{gathered}\xymatrix{
\bullet \ar@{-}[r]\ar@{-}[d]&\bullet \ar@{-}[r]\ar@{-}[d]\ar@{..}[dl] &\bullet \ar@{-}[d]\ar@{..}[dl]&\\
\bullet \ar@{-}[r]\ar@{-}[d]\ar@{..}[dr]&\bullet \ar@{-}[r]\ar@{-}[d]\ar@{..}[dr]&\bullet \ar@{-}[r]\ar@{-}[d]\ar@{..}[dr]&\bullet \ar@{-}[r]\ar@{-}[d]\ar@{..}[dr]&\bullet \ar@{-}[r]\ar@{-}[d]\ar@{..}[dr]&\bullet \ar@{-}[d]\\
\bullet \ar@{-}[r]&\bullet \ar@{-}[r]&\bullet \ar@{-}[r]&\bullet \ar@{-}[r]&\bullet \ar@{-}[r]&\bullet\\
}\end{gathered}$}\\\hline
$\begin{matrix}\\ Q_{2,0} \\ \phantom{a}\end{matrix}$ {\footnotesize (14)} & $\begin{matrix}\\ x^3 +y z^2 + x y^4 \\ \phantom{a}\end{matrix}$ & {\scriptsize$\begin{gathered}\xymatrix{\bullet\ar@{-}[r]\ar@{-}@/_/[dd]&\bullet \ar@{..}[ddl]\ar@{-}@/_/[dd]&\\
\bullet \ar@{-}[r]\ar@{-}[d]&\bullet \ar@{-}[d]\ar@{..}[dl]& \\
\bullet \ar@{-}[r]\ar@{-}[d]\ar@{..}[dr]&\bullet \ar@{-}[r]\ar@{-}[d]\ar@{..}[dr]&\bullet \ar@{-}[r]\ar@{-}[d]\ar@{..}[dr]&\bullet \ar@{-}[r]\ar@{-}[d]\ar@{..}[dr]&\bullet\ar@{-}[d] \\
\bullet \ar@{-}[r]&\bullet \ar@{-}[r]&\bullet \ar@{-}[r]&\bullet \ar@{-}[r]&\bullet\\
}\end{gathered}$}\\\hline
$\begin{matrix}\\ S_{1,0} \\ \phantom{a}\end{matrix}$ {\footnotesize (14)}& $\begin{matrix}\\ x^2 z + y z^2 + y^5 \\ \phantom{a}\end{matrix}$ &{\scriptsize$\begin{gathered}\xymatrix{\bullet\ar@{-}[r]\ar@{-}@/_/[dd]&\bullet \ar@{..}[ddl]\ar@{-}@/_/[dd]&\\
\bullet \ar@{-}[r]\ar@{-}[d]&\bullet \ar@{-}[d]\ar@{..}[dl]\ar@{-}[r]&\bullet \ar@{-}[d]\ar@{..}[dl]\ar@{-}[r]&\bullet \ar@{-}[d]\ar@{..}[dl]&\\
\bullet \ar@{-}[r]\ar@{-}[d]\ar@{..}[dr]&\bullet \ar@{-}[r]\ar@{-}[d]\ar@{..}[dr]&\bullet \ar@{-}[r]\ar@{-}[d]\ar@{..}[dr]&\bullet \ar@{-}[d] \\
\bullet \ar@{-}[r]&\bullet \ar@{-}[r]&\bullet \ar@{-}[r]&\bullet\\
}\end{gathered}$}\\\hline
$\begin{matrix}\\ U_{1,0}  \\ \phantom{a}\end{matrix}$ {\footnotesize (14)} & $\begin{matrix}\\ x^3 + x z^2 + x y^3 \\ \phantom{a}\end{matrix}$ & {\scriptsize$\begin{gathered}\xymatrix{
\bullet\ar@{-}[r]\ar@{-}@/_/[dd]&\bullet \ar@{..}[ddl]\ar@{-}@/_/[dd]\ar@{-}[r]&\bullet \ar@{..}[ddl]\ar@{-}@/_/[dd]&\\
\bullet \ar@{-}[r]\ar@{-}[d]&\bullet \ar@{-}[d]\ar@{..}[dl]\ar@{-}[r]&\bullet \ar@{-}[d]\ar@{..}[dl]& \\
\bullet \ar@{-}[r]\ar@{-}[d]\ar@{..}[dr]&\bullet \ar@{-}[r]\ar@{-}[d]\ar@{..}[dr]&\bullet \ar@{-}[r]\ar@{-}[d]\ar@{..}[dr]&\bullet\ar@{-}[d] \\
\bullet \ar@{-}[r]&\bullet \ar@{-}[r]&\bullet \ar@{-}[r]&\bullet\\
}\end{gathered}$}\\\hline
\end{tabular}
\end{center}
\end{table}

\begin{table}
\begin{center}
\begin{tabular}{|c|c|c|c|}\hline
 & $q_x, q_y, q_z$ & $\hat{c}$ & $\ell$ \\\hline
$E_{19}$ & 1/3 , 1/7 , 1/2 & 8/7 & 18 \\
$Z_{17}$ & 7/24, 1/8, 1/2 & 7/6 & 10 \\
$Z_{18}$ & 5/17, 2/17, 1/2 & 20/17 & 14 \\
$Z_{19}$ & 8/27, 1/9, 1/2 & 32/27 & 22 \\
$W_{17}$ & 1/4, 3/20, 1/2 & 6/5 & 8 \\
$Q_{16}$ & 1/3, 1/7, 3/7 & 25/21 & 17 \\
$Q_{17}$ & 1/3, 2/15, 13/30 & 6/5 & 12 \\
$Q_{18}$ & 1/3, 1/8, 7/16 & 29/24 & 19 \\
$S_{16}$ & 5/17, 3/17, 7/17 & 21/17 & 13 \\
$S_{17}$ & 7/24, 1/6, 5/12 & 5/4 & 9 \\\hline
$Z_{1,0}$ & 2/7, 1/7, 1/2 & 8/7 & 6 \\
$Q_{2,0}$& 1/3, 1/6, 5/12 & 7/6 & 5 \\
$S_{1,0}$& 3/10, 1/5, 2/5 & 6/5 & 4 \\
$U_{1,0}$& 1/3, 2/9, 1/3 & 11/9 & 9 \\\hline
\end{tabular}
\end{center}
\caption{Chiral charges $q_i$, central charges $\hat{c}$ and orders $\ell$ of the quantum monodromy $\mathbb{M}(q)$ for the quasi--homogeneous bimodal singularities that are not square tensor models. }\label{numerology}
\end{table}

Bimodal singularities are fully classified \cite{AGV, EMS}: they are organized in 8 infinite series and 14 exceptional families. All 14 exceptional families have a quasi--homogeneous point in their moduli and, in between the 8 infinite series, there are 6 families that admit one. With an abuse of language, we will refer to this set as to the set of quasi--homogeneous bimodal singularities. The quasi--homogeneous potentials, $W(x,y,z)$, that corresponds to the elements of this set lead to non--degenerate  $2d$  $\cn=(2,2)$ Landau--Ginzburg superconformal field theories that have central charge $\hat{c}<2$, therefore, according to \cite{GVW,SV}, these singularities are all at finite distance in Calabi--Yau moduli space: the local CY 3--fold $\mathscr{H}$, given by the hypersurface in $\mathbb{C}^4$
\begin{equation}\label{IRCY}
\mathscr{H} \colon W(x,y,z) + u^2 + \textrm{ lower terms} = 0.
\end{equation}
is a good candidate for the compactification of type IIB superstring that leads to the engineering of an $\cn = 2$ superconformal $4d$ theory.

In between the theories obtained engineering Type IIB superstring on bimodal singularities there are the following superconformal square tensor models \cite{CVN}:
\begin{equation}
\begin{gathered}
\begin{tabular}{|c|c|c|cc|c|c|c|}\cline{1-3}\cline{6-8}
$E_{18}$ & $x^3 + y^{10} + z^2$ &$ A_2 \square A_9$ &&& $E_{20}$& $x^3 + y^{11} + z^2 $ &  $A_2 \square A_{10}$\\
$W_{18}$& $x^4 + y^7 + z^2 $ & $A_3 \square A_6$ &&& $U_{16}$ & $x^3 + xz^2 + y^5$ & $D_4 \square A_4$\\
$J_{3,0}$ & $x^3 + y^9 + z^2 $ & $A_2 \square A_8$ &&& $W_{1,0} $&$ x^4 + y^6 + z^2 $ & $A_3 \square A_5$\\\cline{1-3}\cline{6-8}
\end{tabular}
\end{gathered}
\end{equation}

In the present work we will focus on the bimodal singularities that are not in this list --- see tables \ref{ArnB1}, \ref{ArnB2}, \ref{numerology}. We remark that also the even rank elements of the $Q$ series are square tensor models, indeed,
\begin{equation}\label{Qs}
\begin{gathered}
\begin{tabular}{|c|c|c|cc|c|c|c|}\cline{1-3}\cline{6-8}
$Q_{10}$ &$ x^2 z + y^3 + z^4$ & $A_2 \square D_5 $&&& $Q_{12}$&$ x^2 z + y^3 + z^5 $ &$ A_2 \square D_6$\\
$Q_{16}$ &$ x^3 + yz^2 + y^7 $&$ A_2 \square D_8$&&&$ Q_{18}$&$ x^3 + yz^2 + y^8 $ &$ A_2 \square D_9$\\\cline{1-3}\cline{6-8}
\end{tabular}
\end{gathered}
\end{equation}

we will discuss this point in \S \ref{2dWC}.

\section{$\cn = 2$ superconformal models}

\subsection{Modality and completeness}\label{ModComp}

\begin{table}
\begin{center}
\begin{tabular}{|cc||cc|}\hline
$Z_{1,0}$&$x^2y^3(1), xy^6(\frac{8}{7})$&$Q_{2,0}$&$x^2y^2(1), x^2y^3(\frac{7}{6})$\\\hline
$S_{1,0}$&$zy^3(1), zy^4(\frac{6}{5})$&$U_{1,0}$&$zy^3(1), zy^4(\frac{11}{9})$\\\hline
\end{tabular}
\end{center}
\begin{center}
\begin{tabular}{|cc|cc|cc|}\hline
$E_{19}$&$y^{11}(\frac{22}{21}), y^{12}(\frac{8}{7})$&$W_{17}$&$y^7(\frac{21}{20}), y^8(\frac{6}{5})$&&\\\hline
$Z_{17}$&$xy^6(\frac{25}{24}), xy^7(\frac{7}{6}) $&$Z_{18}$&$y^9(\frac{18}{17}), y^{10}(\frac{20}{17})$&$Z_{19}$&$xy^7(\frac{29}{27}), xy^8(\frac{32}{27})$\\\hline
$Q_{16}$&$xy^5(\frac{22}{21}), xy^6(\frac{25}{21})$&$Q_{17}$&$y^8(\frac{16}{15}), y^9(\frac{6}{5})$&$Q_{18}$&$xy^6(\frac{13}{12}), xy^7(\frac{29}{24})$\\\hline
$S_{16}$&$y^6(\frac{18}{17}), y^7(\frac{21}{17})$&$S_{17}$&$zy^4(\frac{13}{12}), zy^5(\frac{5}{4})$&&\\\hline
\end{tabular}
\end{center}
\caption{Primary deformations of dimension $\geq 1$ of the quasi--homogeneous bimodal singularities that are not square tensor models. The number in parenthesis is the dimension of the deformation.}\label{deform}
\end{table}

The Landau--Ginzburg models we are considering here have chiral ring of primary operators \cite{VW,LVW}
\begin{equation}\label{chiralring}
\mathscr{R} \simeq \mathbb{C}[x,y,z] / J_W,
\end{equation}
where $J_W$ is the jacobian ideal of $W$, \emph{i.e.} the ideal of $\mathbb{C}[x,y,z]$ generated by the partials $\partial_i W$. The theories being non--degenerate, the ring is finite--dimensional as a $\mathbb{C}$--algebra and its dimension, $\mu$, is called the Milnor number (or multiplicity) of the singularity $W(x,y,z)$,
\begin{equation}\label{Milnornumber}
\mu \equiv \text{dim}_{\mathbb{C}} \mathscr{R}.
\end{equation}
This number equals the Witten--index $\text{tr}(-)^F$ of the theory, by the spectral flow isomorphism \cite{LVW,C91}. Moreover, by $2d/4d$ correspondence $\mu$ equals the rank of the charge lattice $\Gamma$ of the 4d theory. $\text{tr}(-)^F$ jumps at the infrared fixed point obtained by  perturbing the theory away from criticality with primary relevant perturbations: this corresponds to the fact that taking this infrared limit we are projecting on a proper subalgebra of $\mathscr{R}$. \emph{Modality} is the cardinality of the set of perturbations that generate renormalization flows  asymptotically preserving the Witten--index \cite{Gabrielov3}. From this definition, it follows that it can be computed as the number of marginal and irrelevant primary operators in a monomial basis of the chiral ring\footnote{A result known in the mathematical litterature as the fact that inner modality and modality coincide \cite{EMS, GabrielovKushnirenko}.}.

\medskip

The $4d$ theories obtained by geometric engineering Type IIB on $\mathscr{H}$ are \emph{non}--complete quiver quantum field theories in the sense of \cite{CV11}. This means that \cite{CDZ} the image in $\mathbb{C}^{\mu}$ of the domain $\mathcal{D}$ in parameter space that corresponds to consistent QFT's through the holomorphic map $\varpi : \mathcal{D} \rightarrow \mathbb{C}^{\mu}$ defined by the central charge function has nonzero codimension. Variations of the central charge $Z_i \rightarrow Z_i + \d Z_i$ correspond to infinitesimal deformations of the periods of the holomorphic top form $\Omega$ associated to deformations $\d t_{\a}$ of the complex structure of $\mathscr{H}$ of the form
\begin{equation}
W(x,y,z) + u^2 + \sum_{\a} \d t_{\a} \phi_{\a} = 0
\end{equation}
where $\{ \phi_{\a} \} $ is a basis of chiral primaries for the chiral ring. The quantum--obstructed variations $(\d Z_i)_{\textrm{obs}}$ are the ones normal to the physical submanifold $\varpi (\mathcal{D}) \subset \mathbb{C}^{\mu}$. The $2d$ renormalization group allows us to identify such directions. Indeed, the $2d$ theory has infrared conformal fixed point dictated by the Zamolodchikov's $c$--theorem \cite{Zamolod}. The infrared fixed point is stable under perturbation by irrelevant operators. Correspondingly, variations of the periods of $\Omega$ in these directions  of the basis of the Milnor fibration \cite{Milnor} are forbidden physically, since these $2d$ deformations renormalize away. Thus, the unphysical deformations of the theory $(\d Z_i)_{\textrm{obs}}$ are precisely those corresponding to the irrelevant primary perturbations. In computing the BPS spectra, the combinatorics of the quantum clusters is not sensible to this fact: one can compute a mathematically consistent spectrum that corresponds to a BPS chamber that cannot be realized physically due to the above phenomenon ---  this is the root of the \emph{quantum Schottky problem}.

\medskip

In the case of exceptional bimodal singularities, there are two quantum obstructed deformations, therefore the codimension of  $\varpi (\mathcal{D}) \subset \mathbb{C}^{\mu}$ is 2, while for the non--exceptional bimodals, there is only one such deformation, the other modulus being a marginal deformation  ---  see table \ref{deform}. As an example, consider the $Q_{2,0}$ theory. The marginal deformation of it is $x^2 y^2$. This is an operator equivalent, in the chiral ring, to $y^6$. By deforming $Q_{2,0}$ with $y^6$, we obtain the equivalence $Q_{2,0} \sim A_2 \square D_7$.

\subsection{$2d$ wall--crossings and Coxeter--Dynkin graphs}\label{2dWC}

As already mentioned in \cite{CDZ}, Coxeter--Dynkin graphs are \emph{not} unique, depending on the choice of a distinguished homology basis. Equivalent distinguished basis are related by the braid group (Picard--Lefshetz) transformations, \emph{i.e.} by the $2d$ wall--crossing group. Whenever we switch the position of two {\sc susy} vacua, say $| i \rangle$ and $| i+1 \rangle$, in the $W$ plane, we cross a $2d$ wall of marginal stability and this has the effect of a phase transition in the $2d$ BPS spectrum. Accordingly, there is a mapping, $\a_i$, of the strongly distinguished basis $\{ \d_k \}$ into a new one $\{ \d^{\prime} _k \}$ as follows:
\begin{equation}\label{2dWX}
\a_i \colon \begin{cases}
\d^{\prime} _j = \d_j \text{ for } j \neq i, i+1\\
\d^{\prime} _{i+1} = \d_i\\
\d^{\prime} _i = \d_{i+1} + (\d_{i+1} \cdot \d_i ) \, \d_i.
\end{cases}
\end{equation}

If the singularity has Milnor number $\mu$, there are $\mu - 1$ such operations: $\a_i$, $1 \leq i < \mu$. It is easy to check that they satisfy the braid group relations 
\begin{equation}\label{notKeller}
\begin{gathered}
\begin{tabular}{rcl}
$\a_i \a_{i+1} \a_i = \a_{i+1} \a_i \a_{i+1}$ && for $1 \leq i \leq \mu - 2$ \\
$\a_i \a_j = \a_j \a_i $ && for $|i-j| \geq 2$.
\end{tabular}
\end{gathered}
\end{equation}
so that this is a presentation of $B_{\mu}$, the (Artin) braid group with $\mu$ strands\footnote{We stress that the $2d$ wall--crossing group is not, in principle --- according to proposition 9 in I.7.6. of \cite{Bourbaki:Algebra} --- equivalent to the braid group generated by Seidel--Thomas twists, ${\tt Braid}(Q)$ \cite{Keller:2011}, that has generators $\sigma_i$, $i \in Q_0$ and relations:
$$
{\tt Braid}(Q) \colon \begin{cases}
\s_i \s_j = \s_j \s_i \text{ if } i \text{ and } j \text{ are not linked by any arrow}\\
\s_i \s_j \s_i = \s_j \s_i \s_j  \text{ if there is one arrow between } i \text{ and } j\\
none \text{ if there is more than one arrow between }  i \text{ and } j.
\end{cases}
$$

If one considers, say, the Dynkin $A_m$ quiver, ${\tt Braid} (A_m)$ is the $B_{m+1}$ braid \cite{SeidelThomas}, while the $2d$ wall--crossing group is $B_m$. This is the situation whenever the graph is Dynkin: the $2d$ wall--crossing group embeds in ${\tt Braid}(Q)$; the situation gets more complicated whenever the quiver is not a tree.}. Moreover we are allowed by PCT to reverse the orientation of the cycles:
\begin{equation}
r_i \colon \begin{cases}
\d^{\prime} _j = \d_j \text{ for } i \neq j \\
\d^{\prime} _{i} = - \d_i.
\end{cases}
\end{equation}

Consider the $2d$ quantum monodromy, $(S^{-1})^t S$, where
\begin{equation}
S_{ij} = \d_{ij} - \begin{cases}
\d_i \cdot \d_j \text{ for } i < j\\
0 \qquad \text{otherwise}.
\end{cases}
\end{equation}
The spectrum of this operator is physical: 
\begin{equation}\label{2dcheck}
{\tt Eigenvalues}[ (S^{-1})^t S] = \{ \texttt{exp}[2 \pi i ( q_i - \hat{c}/2)] \}
\end{equation}
where $q_i$ are the chiral charges of a basis of chiral primaries at the $\hat{c}$ conformal fixed point. Being physical, the spectrum of the $2d$ quantum monodromy is conserved along the orbits of the $2d$ wall--crossing group\cite{CV92}.

\medskip

Assume that the superpotential of the $\cn = (2,2)$ $2d$ Landau--Ginzburg superconformal theory has nonzero modality, then there are directions along the $2d$ renormalization group flow along which the Witten--index is conserved also asymptotically. The behaviour of the Coxeter--Dynkin graph under these deformations is encoded in the following proposition:
\smallskip

{\bf Proposition}( 1 of \cite{Gabrielov2}): all the irrelevant and marginal deformations of an $\cn = (2,2)$ $2d$ Landau--Ginzburg superconformal theory with $\mu < \infty$ lead to equivalent configurations of vacua and interpolating BPS solitons, where, by equivalent, we mean that they are in the same $2d$ wall--crossing group orbit up to PCT.

\smallskip

This proposition is the key to understand the phenomenon we encountered with the even elements of the $Q$ series: it is just the $2d$ wall--crossing in action. The diagrams we reported in \cite{CDZ} and in table \ref{ArnB1} are referred to an irrelevant deformation, then tuned to zero for consistency of the quantum theory, while the ones from which the square tensor form is explicit are obtained directly from the undeformed theory: being the diagrams in a $\mu$=const. stratum of $Q_{2k}$ they are equivalent.

\medskip

Indeed, for all $k$'s the result is in perfect agreement with \cite{CVN}:
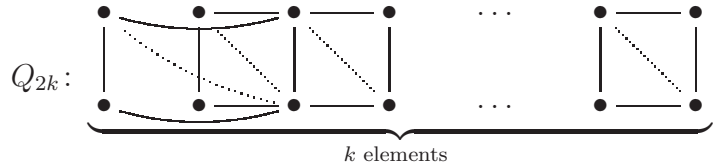
$$Q_{2k} \colon 
\begin{gathered}
\underbrace{\xymatrix{
\bullet\ar@{-}[d]\ar@{-}@/_/[rr]&\bullet\ar@{-}[d]\ar@{-}[r]&\bullet\ar@{..}[dr]\ar@{-}[d]\ar@{-}[r]&\bullet\ar@{-}[d]& \dots &\bullet\ar@{-}[d]\ar@{-}[r]\ar@{..}[dr]&\bullet\ar@{-}[d]\\
\bullet\ar@{-}@/_/[rr]&\bullet\ar@{-}[r]&\bullet\ar@{..}@/^/[ull]\ar@{..}[ul]\ar@{-}[r]&\bullet& \dots &\bullet\ar@{-}[r]&\bullet\\
}}_{k \text{ elements }}
\end{gathered}
$$

\smallskip

We remark that $Q_{14}$ above is just $Q_{2,0}$.

\subsection{$2d$ Renormalization group flows}\label{RGflows}

In \cite{CDZ} both a mathematical and a physical argument in favour of the physical realizability of the BPS chamber in which we will compute the spectra were given. The same argument carries over to the theories obtained from quasi--homogenous bimodal singularities, so let us briefly review the physical (and more stringent) one here. The idea is that the 14 superconformal models which are not already of the form $G\square G^{\prime}$ can be obtained each from an appropriate square tensor model of type $A_n \square G$ by perturbing it with a suitable IR--relevant operator\footnote{IR--relevant at the UV--fixed point described by the $A_n \square G$ theory.} $\phi_{\star}$. The IR--relevant perturbation is chosen in such a way that the corresponding $\cn = 2$ theory will flow in the IR to the given Arnold superconformal theory. 

\medskip

The trivial instances of this RG process are the following theories (we indicate in parenthesis the dimension of $\phi_{\star}$):
\begin{equation}
\begin{gathered}
\begin{tabular}{c}
$A_2 \square A_{10} \colon x^3 + y^{11} + z^2  \xrightarrow{\ xy^7 \ (32/33)\ } E_{19}$\\
$A_3 \square A_7 \colon x^4 + y^8 + z^2  \xrightarrow{\ x^3 y \ (7/8)\ } Z_{17}$\\
$A_3 \square A_6 \colon x^4 + y^7 + z^2 \xrightarrow{\ x y^5 \ (27/28)\ } W_{17}$\\
$A_3 \square A_6 \colon x^4 + y^7 + z^2 \xrightarrow{\ x^3 y \ (25/28)\ } Z_{1,0}$\\
$A_4 \square D_4 \colon x^2 z + x^3 + y^5 \xrightarrow{\ x y^3 \ (14/15)\ } U_{1,0}$.\\
$A_4 \square D_4 \colon x^2 z + z^3 + y^5 \xrightarrow{\ y z^2 \ (13/15)\ } S_{1,0}$\\
\end{tabular}
\end{gathered}
\end{equation}

The theories $Z_{18}, S_{16}$ and the ones of the $Q$ series are better described as the final IR points of RG `cascades'

\begin{equation}
\begin{gathered}
\begin{tabular}{c}
$A_3 \square A_8 \colon x^4 + y^9 + z^2  \xrightarrow{\ x^3 y \ (31/36)\ } Z_{19} \xrightarrow{\ x y^6 \ (26/27)\ } Z_{18} $\\
$A_5 \square D_4 \colon x^2 z + z^3 + y^6 \xrightarrow{\ y z^2 \ (5/6)\ } S_{17} \xrightarrow{\ x y^4 \ (23/24)\ } S_{16}$\\
$A_6 \square D_4 \colon x^3 + z^3 + y^7 \xrightarrow{\ y z^2 \ (17/21)\ } Q_{16} \xrightarrow{\ x y^4 \ (19/21)\ } Q_{2,0}$\\
$A_7 \square D_4 \colon x^3 + z^3 + y^8 \xrightarrow{\ y z^2 \ (19/24)\ } Q_{18} \xrightarrow{\ x y^5 \ (23/24)\ } Q_{17}$.\\
\end{tabular}
\end{gathered}
\end{equation}

As explained in \cite{CDZ}, this RG argument applies directly to the theories at their superconformal point, \emph{i.e.} when all relevant deformations are switched off. We can give volumes to the (special lagrangian) 3--cycles $\gamma_i$ in the third homology group of the Calabi--Yau 3--fold
$$ y^{n+1} + W_{G}(x,z) + u^2 = 0,$$
by the primary deformation of this singularity. The D3--branes that wrap around these 3--cycles, therefore, get central charges
$$ Z(\gamma_i) = \int_{\gamma_i} \Omega \, ,$$
becoming the BPS particles of the massive deformation of the corresponding 4 dimensional superconformal theory. If now we deform this theory with $\phi_{\star}$ and with chiral primaries of dimension $q$ less than $q(\phi_{\star})$, along the 2 dimensional RG flow in the infrared some of the above 3--cycles start increasing their volume, that becames bigger and bigger the more close we are to the IR fixed point. Accordingly, the corresponding BPS particle masses increases. Thus, these particles decouples and are absent from the BPS spectrum of the 4 dimensional theory obtained by engineering type IIB on the corresponding primary deformation of $\mathscr{H}$, the Calabi--Yau 3--fold associated to the IR theory. Obviously, the two spectra are comparable if, along the flow line of the 2 dimensional RG, we do not cross any wall of marginal stability in 4 dimensions. Since checking this fact may be difficult in practice\footnote{One has to check the existence of a `tuning' of the complex deformation $\lambda \phi_{\star}$ such that it gives rise to a path that avoids the wall--crossings, while keeping control of the possible mixing between the conserved quantum currents.}, we use the above idea in the weak sense of \cite{CDZ}: whenever a (mathematically correct) BPS spectrum can be naturally interpreted as obtained from a physically realized one by the decoupling of some heavy states along the 2 dimensional RG flow above described, we take this fact just as a circumstantial evidence for the corresponding BPS chamber to be physically realizable.

\subsection{Flavor charges}

An important invariant of the theory, \emph{i.e.} a quantity that remains uniformly constant over $\mathcal{D}$, is the number of flavor charges, $n_f$, the dimension of the Cartan subalgebra of the flavor symmetry group. At a generic point of $\mathcal{D}$ the theory has flavor group $U(1)^{n_f}$, while at certain points of parameter space this symmetry can enhance to a non--Abelian one $G_f$. From the point of view of BPS quiver theory, $n_f$ is just the number of zero eigenvalues of the exchange matrix of the quiver, indeed, a charge $\gamma_f$ is flavor if and only if
\begin{equation}\label{flatt}
 \langle \gamma , \gamma_f \rangle_D = 0 \qquad \forall \, \gamma \in \Gamma
 \end{equation}
where $\Gamma$ is the charge lattice of the theory. In particular, $n_f = \textrm{rank } \Gamma \, \textrm{ mod } 2$. A general consequence of $2d/4d$ correspondence \cite{CVN, CV11, CDZ} is that $n_f$ is equal to the number of chiral primaries of dimension $\hat{c}/2$. As explained in \cite{CDZ}, as far as the non--exceptional theories one can read off this number from table 2 of ref. \cite{lenzing}, it is just the number of $\Phi_2$ factors in the factorization of the characteristic polynomial of the strong monodromy $H$ in cyclotomic polynomials\footnote{Although, there is a misprint there: the correct flavor charge of $Z_{1,0}$ is 3.}. As far as the other singularities, the degeneracies of the chiral ring elements are captured by the Poincar\'e polynomial:
\begin{equation}
\sum_\a t^{q_{\a}} = \prod_i \frac{(1- t^{1-q_i})}{1-t^{q_i}}
\end{equation}

where the $q_i$ are the charges of table \ref{numerology} and the sum is over all the elements of a monomial basis of the chiral ring. Expanding the RHS, $n_f$ is the (positive or zero) \emph{integer} multiplying the coefficient $t^{\hat{c}/2}$ of the serie. So,
\begin{equation}\label{flavorch}
\begin{gathered}
n_f = \begin{cases}
3 \textrm{ for } Z_{1,0}\\
2 \textrm{ for } Q_{2,0}, S_{1,0}, J_{3,0}, Z_{18} \\
1 \textrm{ for odd rank exceptionals and } W_{1,0}\\
0 \textrm{ otherwise.}
\end{cases}
\end{gathered}
\end{equation} 

\subsection{A remark about weak coupling}

Let us end this section with a remark about the consequences of the above analysis on the weak coupling limit of these theories, although we will not discuss it in the present paper. Consider the possibility that one of the above theories admits in some corner of its parameter space the structure of a $G$ SYM weakly coupled to some other sector. Assume, momentarily, that such a description is purely lagrangian. The dimension of the parameter space, in this case, can be computed as
\begin{equation}
\begin{gathered}
\text{dim }(\mathcal{D}) = \# (\text{gauge couplings}) + \text{ dim }(\text{Coulomb branch}) + \#(\text{masses})\\
= \# (\text{simple factors of G}) + \text{rk}(G) + n_f
\end{gathered}
\end{equation}
By the $4d/2d$ correspondence, $\mu$ is equal to the rank of the charge lattice $\Gamma$ that, in this case, is given by \cite{CV11}:
\begin{equation}\label{rkGamma}
\text{rk}(\Gamma) = 2 \text{ rk}(G) + n_f.
\end{equation}
Thus
\begin{equation}\label{groupmagic}
\mu - \text{dim }(\mathcal{D}) = \text{rk}(G) -  \# (\text{simple factors of G}) = \text{codim} (\mathcal{D}).
\end{equation}
This equality holds in a \emph{lagrangian} corner of the parameter space: for a non--lagrangian one we expect that it becames an inequality,
\begin{equation}
\text{rk}(G) -  \# (\text{simple factors of G}) \leq \text{codim} (\mathcal{D}),
\end{equation}
since there could be more complicated mechanisms that lead to forbidden directions. From \S 3.1 --- see table 4 --- we are able to compute the codimension of $\mathcal{D}$ in $\mathbb{C}^{\mu}$ 
\begin{equation*}
\text{codim} (\mathcal{D}) = \begin{cases} 2 \text{  for exceptional bimodals;}\\
1 \text{  otherwise.}
\end{cases}
\end{equation*}
Therefore, if one of the theories we are considering has the structure of a $G$ SYM weakly coupled to some other sector that maybe non--lagrangian, just counting dimensions, we are able to constrain the possible gauge groups $G$: for non-exceptional bimodals the possibilities are
\begin{equation}
SU(2)^k,\quad SU(2)^k\times SU(3),\quad SU(2)^k\times SO(5),\quad SU(2)^k\times G_2,
\end{equation}
while for exceptional bimodals we have the above cases and the following ones
\begin{equation}
\begin{gathered}
SU(2)^k\times SU(3)\times SO(5),\quad SU(2)^k\times SU(3)\times G_2, \\
SU(2)^k\times SO(5)\times G_2, \quad SU(2)^k\times SU(4),\\
SU(2)^k\times SO(6),\quad SU(2)^k\times SO(7)
\end{gathered}
\end{equation}
for some $k \in \mathbb{N}$. Since $\mu = \text{rk}(\Gamma)$, the possible number of $SU(2)$ factors appearing here is constrained via \eqref{rkGamma} by the Witten--index of the corresponding 2d theory.


\section{BPS spectra at strong coupling}

\subsection{Quivers and potentials}\label{CDZMethod}

In ref.\cite{CDZ}, by the $2d/4d$ correspondence, a method for obtaining a $4d$ BPS quiver with potential from a Coxeter--Dynkin graph of the corresponding $2d$ SCFT was outlined. The method carries over in all cases for which the resulting basic algebra of step 2, $\mathscr{A}$, is such that $\texttt{gl.dim.}\mathscr{A} \leq 2$. It consists of the following four steps:

\smallskip

\emph{Step 0:} find an appropriate $2d$ configuration of vacua and BPS solitons in the $2d$ wall--crossing group orbit of the theory.

\smallskip

\emph{Step 1:} choose an orientation of the solid edges of the Coxeter--Dynkin graph such that the dashed edges make sense as relations in the path algebra of the quiver $Q$ so obtained.

\smallskip

\emph{Step 2:} the dashed edges are interpreted as generating an ideal $J$ in the path algebra $\mathbb{C} Q$. $\mathscr{A}$ is the basic algebra $\mathbb{C}Q / J$.

\smallskip

\emph{Step 3:} interpret the ideal $J$ as the Jacobian ideal of a $1d$ superpotential $\mathcal{W}$. Add to $Q$ the arrows that allows you to interpret the relations as $F$--term flatness conditions. This way the completed quiver $\widetilde{Q}$ is obtained. The relevant algebra --- \emph{i.e.} the one whose stable representations are related to BPS spectra --- is the 3--CY completion of $\mathscr{A}$, $\Pi_3 (\mathscr{A}) \simeq  \mathbb{C} \widetilde{Q} / \partial \mathcal{W} $.

\smallskip

Let us notice that this superpotential is interpreted as describing the supersymmetric quantum mechanics that governs the dynamics of the worldline of the D--brane system used in the engineering of the theory. With the above method we have obtained quivers with potentials ($\widetilde{Q}$,$\mathcal{W}$) for all the theories in the present paper. For each theory, starting from this representative of the quiver with potential, one can easily obtain, by repeated mutations, a \emph{square form} representative --- \emph{i.e.} the one obtained by eliminating all the dashed arrows from the Coxeter--Dynkin diagram and orienting all the squares; the superpotential of this quiver is given by the traces of the cycles corresponding to the oriented squares.

\subsection{Finite BPS chambers}\label{BPSspectra}

All the square form representatives of the BPS quivers of the $\cn = 2$ theories that we are considering, admit decompositions in complete families of $ADE$ Dynkin subquivers \cite{CDZ} $\{ G_a \}_{a \in A}$, this means that the charge lattice $\Gamma$ has an isomorphism with the following direct sum of root lattices of Lie algebras:
\begin{equation}
\Gamma \simeq \bigoplus_{a \in A} \Gamma_{G_a}.
\end{equation}
To these decompositions, moreover, there correspond Weyl--factorized sink--sequences of Coxeter--type and, therefore, there are some algebraically obvious finite BPS chambers. In these chambers, the BPS spectra consist of one hypermultiplet per charge vector of the direct--sum form
\begin{equation}
0\oplus ... \oplus \alpha^{(a)} \oplus 0 \oplus ... \oplus 0, \qquad \alpha^{(a)} \in \Delta_+(G_a).
\end{equation}
having only one non--zero component (equal to a positive root of the corresponding Lie algebra $G_a$). In these cases the consistency of the mass spectrum follows from comparison with the (obviously consistent) mass spectrum of the $G_a$--type Argyres--Douglas models in the maximal chamber \cite{CVN}. Our result are the following\footnote{The notation $(..., G \times N, ... )$ means that the Dynkin graph $G$ appears $N$ times in the family.}:
\begin{itemize}
\item $ { \bf E_{19} }$ : this theory is a one--point extension of $A_2 \square A_9$. We have four algebraically trivial finite chambers:
\begin{equation}\label{E19ch}
\begin{tabular}{c|c}
$( A_2 \times 8 , A_3 )$ & $(A_2 \times 9, A_1)$\\\hline
$(A_{10} , A_9)$ & $(A_9 \times 2, A_1) $
\end{tabular}
\end{equation} 

\item ${ \bf Z_{17}}$ : we have two algebraically trivial finite chambers:
\begin{equation}\label{Z17ch}
\begin{tabular}{c|c}
$( A_3, A_7 , A_7 )$ & $(A_3 \times 3, A_2 \times 4)$\\
\end{tabular}
\end{equation} 

\item ${\bf Z_{18}}$ : this is the one point extension of the previous one:
\begin{equation}\label{Z18ch}
\begin{tabular}{c|c}
$( A_3, A_7 , A_8)$ & $( A_3, A_7 , A_7, A_1)$\\\hline
$(A_3 \times 3, A_2 \times 3, A_3)$ & $(A_3 \times 3, A_2 \times 4, A_1)$
\end{tabular}
\end{equation} 

\item $ {\bf Q_{2k}} $ : the canonical chambers of $A_2 \square D_k$ --- see eq. \eqref{Qs} --- and the following two algebraically trivial finite chambers:
\begin{align}
&{\bf Q_{2,0}} \colon \begin{tabular}{c|c}
$( D_4 \times 2, A_2 \times 3)$ & $(A_2 \times 2, A_5 \times 2)$\\
\end{tabular}\\
&{\bf Q_{16}} \colon \begin{tabular}{c|c}
$( D_4 \times 2, A_2 \times 4)$ & $(A_2 \times 2, A_6 \times 2)$ \label{Q16ch}\\
\end{tabular}\\
&{\bf Q_{18}} \colon \begin{tabular}{c|c}
$( D_4 \times 2, A_2 \times 5)$ & $(A_2 \times 2, A_7 \times 2)$\label{Q18ch}\\
\end{tabular}
\end{align} 

\item ${\bf Q_{17}}$ : this is a one point extension of $Q_{16}$:
\begin{equation}\label{Q17ch}
\begin{tabular}{c|c}
$( D_4 \times 2, A_2 \times 4,A_1)$ & $(A_2 \times 2, A_6 \times 2,A_1)$\\\hline
$( D_4 \times 2, A_2 \times 3,A_3)$ & $(A_2 \times 2, A_6, A_7)$
\end{tabular}
\end{equation} 

\item For all the others we have two algebraically trivial finite chambers:
\begin{align}
&{\bf Z_{19}} \colon
\begin{tabular}{c|c}
$( A_3, A_8 , A_8 )$ & $(A_3 \times 3, A_2 \times 5)$\\
\end{tabular}\\
&{\bf W_{17}} \colon
\begin{tabular}{c|c}
$( A_5, A_6 , A_6 )$ & $(A_3 \times 5, A_2 )$\label{W17ch} \\
\end{tabular}\\
&{\bf S_{16}} \colon 
\begin{tabular}{c|c}
$( D_4 \times 2, A_3 \times 2, A_2)$ & $(A_2, A_4, A_5 \times 2)$\\
\end{tabular}\\\label{S17ch}
&{\bf S_{17}} \colon 
\begin{tabular}{c|c}
$( D_4 \times 2, A_3 \times 3)$ & $(A_2, A_5 \times 3)$\\
\end{tabular}\\
&{\bf Z_{1,0}} \colon
\begin{tabular}{c|c}
$( A_3 , A_6 \times 2)$ & $(A_3 \times 3, A_2 \times 3)$\\
\end{tabular}\\\label{S10ch}
&{\bf S_{1,0}} \colon \begin{tabular}{c|c}
$( D_4 \times 2, A_3 \times 2)$ & $(A_2, A_4 \times 3)$\\
\end{tabular}\\
&{\bf U_{1,0}} \colon
\begin{tabular}{c|c}
$( D_4 \times 3, A_2)$ & $(A_3 \times 2, A_4 \times 2)$\\
\end{tabular}
\end{align}
\end{itemize}

We stress that all these finite BPS chambers have natural physical interpretations as the decoupling of some heavy hypermultiplet from the physical BPS spectrum of a canonical chamber of a square tensor model \cite{CVN}, as we showed in section \S \ref{RGflows}. As already remarked in \cite{CDZ}, we stress that in general it is difficult to understand whether a given chamber is physical or not, even at the heuristic level. This is one of the reasons why we have limited ourself to the study of those chambers that are canonically related to the analysis of \cite{CVN}.

\section{More periodic $Y$--systems}
According to our analysis we are predicting the existence of 11 new periodic $Y$--systems that can be straightforwardly generated with the help of the Keller's mutation applet \cite{applet} using the Weyl--factorized sequences that corresponds to the BPS chambers we listed in \S \ref{BPSspectra} --- see appendix \ref{sequences}. Indeed, BPS spectroscopy provides expressions for the quantum monodromy $\mathbb{M}(q)$. The action of $\mathbb{M}(q)$ on the quantum torus algebra $\mathbb{T}_Q$ is specified by its action on the set of generators $\{Y_i\}_{i \in Q_0}$, where $Q_0$ denotes the set of nodes of $Q$,
\begin{equation}\label{Mon}
Y_i \rightarrow Y_i^{\prime} \equiv \text{Ad}(\mathbb{M}(q)^{-1})Y_i \equiv N[R_i (Y_j)],
\end{equation}
where $N[...]$ is the normal--ordering \cite{CVN}. The classical limit of \eqref{Mon}, is a rational map $R\colon Y_i \rightarrow R_i (Y_j)$, the iteration of which is the $Y$--system:
\begin{equation}
Y_i(s+1) = R_i (Y_j(s)), \qquad s\in \mathbb{Z}.
\end{equation}
The  $Y$--systems so obtained are \emph{periodic} since, 
\begin{equation}
\text{Ad} \big[ \mathbb{M}(q)^{\ell} \big] = \text{ Id} \Longleftrightarrow Y_j(s+\ell) = Y_j(s), \forall \,  j \in Q_0, s \in \mathbb{Z},
\end{equation}
and string theory predicts \cite{CDZ,CVN} the values of the orders $\ell$ for the models we have studied in this paper --- see the list in table \ref{numerology}. Moreover, we have checked the RHS with the help of the computer procedure described in \cite{CDZ}. From our analysis, the periodic $Y$--systems associated to the $Q$--series of unimodal and bimodal singularities should be interpreted as new forms of the $A_2 \square D_k$ ones. 

\medskip

As remarked in \cite{CDZ}, it should be possible to give an interpretation of these new periodic $Y$--systems in terms of exactly solvable $2d$ theories in analogy with the $(G,G^{\prime})$ ones \cite{Yrefs}.

\section*{Aknowledgments}
The author wants to thank Sergio Cecotti for his enlightening teachings. Moreover, we acknowledge Wolfgang Ebeling for having cleared up to us a point of fundamental importance for our analysis.

\appendix

\section{Sink--sequences}\label{sequences}

In this appendix we discuss some selected examples of the Weyl--factorized sequences we have used to generate the BPS spectra of \S \ref{BPSspectra}. We use the conventions of \cite{ASS}.

\begin{itemize}
\item ${\bf E_{19}}$

\smallskip

This model is a one point extension of $A_2 \square A_9$. We will discuss all the chambers of it, since it is the simplest one point extension in between the models we are studying in this paper and the other one point extensions behaves similarly.

The square form of the $Q_{E_{19}}$ quiver is 
$$ Q_{E_{19}} \colon \begin{gathered}
 \begin{xy} 0;<0.6pt,0pt>:<0pt,-0.6pt>:: 
(0,50) *+{1} ="0",
(0,0) *+{2} ="1",
(50,50) *+{3} ="2",
(50,0) *+{4} ="3",
(100,50) *+{5} ="4",
(100,0) *+{6} ="5",
(150,50) *+{7} ="6",
(150,0) *+{8} ="7",
(200,50) *+{9} ="8",
(200,0) *+{10} ="9",
(250,50) *+{11} ="10",
(250,0) *+{12} ="11",
(300,50) *+{13} ="12",
(300,0) *+{14} ="13",
(350,50) *+{15} ="14",
(350,0) *+{16} ="15",
(400,50) *+{17} ="16",
(400,0) *+{18} ="17",
(450,50) *+{19} ="18",
"1", {\ar"0"},
"0", {\ar"2"},
"3", {\ar"1"},
"2", {\ar"3"},
"4", {\ar"2"},
"3", {\ar"5"},
"5", {\ar"4"},
"4", {\ar"6"},
"7", {\ar"5"},
"6", {\ar"7"},
"8", {\ar"6"},
"7", {\ar"9"},
"9", {\ar"8"},
"8", {\ar"10"},
"11", {\ar"9"},
"10", {\ar"11"},
"12", {\ar"10"},
"11", {\ar"13"},
"13", {\ar"12"},
"12", {\ar"14"},
"15", {\ar"13"},
"14", {\ar"15"},
"16", {\ar"14"},
"15", {\ar"17"},
"17", {\ar"16"},
"16", {\ar"18"},
\end{xy} 
\end{gathered}$$

The original form of $Q_{E_{19}}$ is related to this one by the following sequence of mutations: (1 3 5 7 9 11 13 15 2 4 6 8 10 12 14 1 3 5 7 9 11 2 4 6 8 10 1 3 5 7 2 4 6 1 3 2$)^{-1}$.

\medskip

From $Q_{E_{19}}$ one can easily obtain two of the four chambers \eqref{E19ch}:
\begin{itemize}
\item[(a)] $(A_{10},A_9)$ chamber:
\begin{itemize}
\item complete family: $A_{10}$: odd nodes; $A_{9}$: even nodes 
\item Weyl--factorized sequence: 2 6 10 14 18 3 7 11 15 19 4 8 12 16 1 5 9 13 17 2 6 10 14 3 7 11 15 19 4 8 12 18 1 5 9 13 17 2 6 10 16 3 7 11 15 19 4 8 14 18 1 5 9 13 17 2 6 12 16 3 7 11 15 19 4 10 14 18 1 5 9 13 17 2 8 12 16 3 7 11 15 19 6 10 14 18 1 5 9 13 17 4 8 12 16 3 7 11 15 19
\item permutation:
$$ \boldsymbol{m}_{\Lambda} (Q_{E_{19}})=
\begin{gathered}
\begin{xy} 0;<0.6pt,0pt>:<0pt,-0.6pt>:: 
(0,50) *+{19} ="0",
(0,0) *+{18} ="1",
(50,50) *+{17} ="2",
(50,0) *+{16} ="3",
(100,50) *+{15} ="4",
(100,0) *+{14} ="5",
(150,50) *+{13} ="6",
(150,0) *+{12} ="7",
(200,50) *+{11} ="8",
(200,0) *+{10} ="9",
(250,50) *+{9} ="10",
(250,0) *+{8} ="11",
(300,50) *+{7} ="12",
(300,0) *+{6} ="13",
(350,50) *+{5} ="14",
(350,0) *+{4} ="15",
(400,50) *+{3} ="16",
(400,0) *+{2} ="17",
(450,50) *+{1} ="18",
"1", {\ar"0"},
"0", {\ar"2"},
"3", {\ar"1"},
"2", {\ar"3"},
"4", {\ar"2"},
"3", {\ar"5"},
"5", {\ar"4"},
"4", {\ar"6"},
"7", {\ar"5"},
"6", {\ar"7"},
"8", {\ar"6"},
"7", {\ar"9"},
"9", {\ar"8"},
"8", {\ar"10"},
"11", {\ar"9"},
"10", {\ar"11"},
"12", {\ar"10"},
"11", {\ar"13"},
"13", {\ar"12"},
"12", {\ar"14"},
"15", {\ar"13"},
"14", {\ar"15"},
"16", {\ar"14"},
"15", {\ar"17"},
"17", {\ar"16"},
"16", {\ar"18"},
\end{xy} 
\end{gathered}$$
\item type: $(A_{10}, c^{-5} s_{10} s_8 s_6 s_4 s_2 : A_9, c^5)$
\end{itemize}

\item[(b)] $(A_2 \times 9, A_1)$ chamber:
\begin{itemize}
\item complete family: $A_2 = \{1, 2 \} + 2k, k=0,...,8$, $A_1 = \{ 19 \}$.
\item Weyl--factorized sequence: 1 4 5 8 9 12 13 16 17 2 3 6 7 10 11 14 15 18 1 4 5 8 9 12 13 16 17 19
\item permutation: $$ \boldsymbol{m}_{\Lambda} (Q_{E_{19}})=
\begin{gathered}
 \begin{xy} 0;<0.6pt,0pt>:<0pt,-0.6pt>:: 
(0,50) *+{2} ="0",
(0,0) *+{1} ="1",
(50,50) *+{4} ="2",
(50,0) *+{3} ="3",
(100,50) *+{6} ="4",
(100,0) *+{5} ="5",
(150,50) *+{8} ="6",
(150,0) *+{7} ="7",
(200,50) *+{10} ="8",
(200,0) *+{9} ="9",
(250,50) *+{12} ="10",
(250,0) *+{11} ="11",
(300,50) *+{14} ="12",
(300,0) *+{13} ="13",
(350,50) *+{16} ="14",
(350,0) *+{15} ="15",
(400,50) *+{18} ="16",
(400,0) *+{17} ="17",
(450,50) *+{19} ="18",
"1", {\ar"0"},
"0", {\ar"2"},
"3", {\ar"1"},
"2", {\ar"3"},
"4", {\ar"2"},
"3", {\ar"5"},
"5", {\ar"4"},
"4", {\ar"6"},
"7", {\ar"5"},
"6", {\ar"7"},
"8", {\ar"6"},
"7", {\ar"9"},
"9", {\ar"8"},
"8", {\ar"10"},
"11", {\ar"9"},
"10", {\ar"11"},
"12", {\ar"10"},
"11", {\ar"13"},
"13", {\ar"12"},
"12", {\ar"14"},
"15", {\ar"13"},
"14", {\ar"15"},
"16", {\ar"14"},
"15", {\ar"17"},
"17", {\ar"16"},
"16", {\ar"18"},
\end{xy}
\end{gathered}$$
\item type: $A_2, s_1s_2s_1$ for the even $k$,  $A_2, s_2 s_1 s_2 $ for the odd $k$, $A_1, s$. 
\end{itemize}
\end{itemize}

As far as the other two chambers one has to consider the quiver $\mu_{19} (Q_{E_{19}})$:

\begin{itemize}
\item[(a)] $(A_9 \times 2, A_1)$ chamber:
\begin{itemize}
\item complete family: $A_9 = \{1, 3, 5, 7, 9, 11, 13, 15, 17 \} + k, k = 0,1$, $A_1 = \{ 19 \}$
\item Weyl--factorized sequence: 3 7 11 15 2 6 10 14 18 19 1 5 9 13 17 4 8 12 16 3 7 11 15 2 6 10 14 18 1 5 9 13 17 4 8 12 16 3 7 11 15 2 6 10 14 18 1 5 9 13 17 4 8 12 16 3 7 11 15 2 6 10 14 18 1 5 9 13 17 4 8 12 16 3 7 11 15 2 6 10 14 18 1 5 9 13 17 4 8 12 16
\item permutation: $$ \boldsymbol{m}_{\Lambda} (Q_{E_{19}})=
\begin{gathered}
\begin{xy} 0;<0.6pt,0pt>:<0pt,-0.6pt>:: 
(0,50) *+{17} ="0",
(0,0) *+{18} ="1",
(50,50) *+{15} ="2",
(50,0) *+{16} ="3",
(100,50) *+{13} ="4",
(100,0) *+{14} ="5",
(150,50) *+{11} ="6",
(150,0) *+{12} ="7",
(200,50) *+{9} ="8",
(200,0) *+{10} ="9",
(250,50) *+{7} ="10",
(250,0) *+{8} ="11",
(300,50) *+{5} ="12",
(300,0) *+{6} ="13",
(350,50) *+{3} ="14",
(350,0) *+{4} ="15",
(400,50) *+{1} ="16",
(400,0) *+{2} ="17",
(450,50) *+{19} ="18",
"1", {\ar"0"},
"0", {\ar"2"},
"3", {\ar"1"},
"2", {\ar"3"},
"4", {\ar"2"},
"3", {\ar"5"},
"5", {\ar"4"},
"4", {\ar"6"},
"7", {\ar"5"},
"6", {\ar"7"},
"8", {\ar"6"},
"7", {\ar"9"},
"9", {\ar"8"},
"8", {\ar"10"},
"11", {\ar"9"},
"10", {\ar"11"},
"12", {\ar"10"},
"11", {\ar"13"},
"13", {\ar"12"},
"12", {\ar"14"},
"15", {\ar"13"},
"14", {\ar"15"},
"16", {\ar"14"},
"15", {\ar"17"},
"17", {\ar"16"},
"18", {\ar"16"},
\end{xy}
\end{gathered}$$

\item type: $(A_9, c^{-5}: A_9, c^5: A_1, s)$
\end{itemize}

\item[(b)] $(A_2 \times 8, A_3)$ chamber:
\begin{itemize}
\item complete family: $A_2 = \{1,2\} + 2k, k=0,7$, $A_3 = \{ 17, 18, 19 \}$
\item Weyl--factorized sequence: 1 4 5 8 9 12 13 16 17 2 3 6 7 10 11 14 18 19 15 1 4 5 8 9 12 13 17 16 19 18
\item permutation: $$ \boldsymbol{m}_{\Lambda} (Q_{E_{19}})=
\begin{gathered}
 \begin{xy} 0;<0.6pt,0pt>:<0pt,-0.6pt>:: 
(0,50) *+{2} ="0",
(0,0) *+{1} ="1",
(50,50) *+{4} ="2",
(50,0) *+{3} ="3",
(100,50) *+{6} ="4",
(100,0) *+{5} ="5",
(150,50) *+{8} ="6",
(150,0) *+{7} ="7",
(200,50) *+{10} ="8",
(200,0) *+{9} ="9",
(250,50) *+{12} ="10",
(250,0) *+{11} ="11",
(300,50) *+{14} ="12",
(300,0) *+{13} ="13",
(350,50) *+{16} ="14",
(350,0) *+{15} ="15",
(400,50) *+{17} ="16",
(400,0) *+{19} ="17",
(450,50) *+{18} ="18",
"1", {\ar"0"},
"0", {\ar"2"},
"3", {\ar"1"},
"2", {\ar"3"},
"4", {\ar"2"},
"3", {\ar"5"},
"5", {\ar"4"},
"4", {\ar"6"},
"7", {\ar"5"},
"6", {\ar"7"},
"8", {\ar"6"},
"7", {\ar"9"},
"9", {\ar"8"},
"8", {\ar"10"},
"11", {\ar"9"},
"10", {\ar"11"},
"12", {\ar"10"},
"11", {\ar"13"},
"13", {\ar"12"},
"12", {\ar"14"},
"15", {\ar"13"},
"14", {\ar"15"},
"16", {\ar"14"},
"15", {\ar"17"},
"17", {\ar"16"},
"18", {\ar"16"},
\end{xy} 
\end{gathered}$$
\item type: $A_2, s_1s_2s_1$ for the even $k$,  $A_2, s_2 s_1 s_2 $ for the odd $k$, $A_3, c^{-2}$.
\end{itemize}
\end{itemize}

This is the generic situation for all the one--point extensions: mutating at the extra--node leads two different couples of canonical chambers.

\item ${\bf Z_{17}}$

\smallskip

This will be our example of a theory in the $Z$ family. The  square form of $Q_{Z_{17}}$ is:
$$ Q_{Z_{17}} \colon 
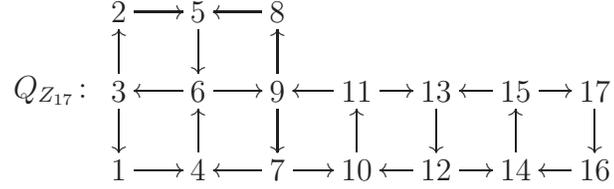
\begin{gathered}
\begin{xy} 0;<0.6pt,0pt>:<0pt,-0.6pt>:: 
(0,100) *+{1} ="0",
(0,0) *+{2} ="1",
(0,50) *+{3} ="2",
(50,100) *+{4} ="3",
(50,0) *+{5} ="4",
(50,50) *+{6} ="5",
(100,100) *+{7} ="6",
(100,0) *+{8} ="7",
(100,50) *+{9} ="8",
(150,100) *+{10} ="9",
(150,50) *+{11} ="10",
(200,100) *+{12} ="11",
(200,50) *+{13} ="12",
(250,100) *+{14} ="13",
(250,50) *+{15} ="14",
(300,100) *+{16} ="15",
(300,50) *+{17} ="16",
"2", {\ar"0"},
"0", {\ar"3"},
"2", {\ar"1"},
"1", {\ar"4"},
"5", {\ar"2"},
"3", {\ar"5"},
"6", {\ar"3"},
"4", {\ar"5"},
"7", {\ar"4"},
"5", {\ar"8"},
"8", {\ar"6"},
"6", {\ar"9"},
"8", {\ar"7"},
"10", {\ar"8"},
"9", {\ar"10"},
"11", {\ar"9"},
"10", {\ar"12"},
"12", {\ar"11"},
"11", {\ar"13"},
"14", {\ar"12"},
"13", {\ar"14"},
"15", {\ar"13"},
"14", {\ar"16"},
"16", {\ar"15"},
\end{xy}
\end{gathered}$$

Related to the original form obtained from its Coxeter--Dynkin diagram of table \ref{ArnB1} by the sequence of mutations: (17 15 13 11 2 1 5 4 3 16 14 12 17 15 16$)^{-1}$. We have two (physical) chambers:

\begin{itemize}
\item[(a)] $(A_7 \times 2, A_3)$ chamber:
\begin{itemize}
\item complete family: $A_7 = \{ 1, 4, 7, 10, 12, 14, 16 \}$ $A_7 = \{ 3, 6, 9, 11, 13, 15, 17 \}$ $A_3 = \{ 2, 5, 8 \}$
\item Weyl--factorized sequence: 4 10 14 3 9 13 17 5 1 7 12 16 6 11 15 2 4 10 14 3 9 13 17 1 7 12 16 6 11 15 4 10 14 3 9 13 17 1 7 12 16 6 11 15 4 10 14 3 9 13 8 17 5 1 7 12 16 6 11 15 2 8
\item permutation:
$$ \boldsymbol{m}_{\Lambda} (Q_{ Z_{17}})=
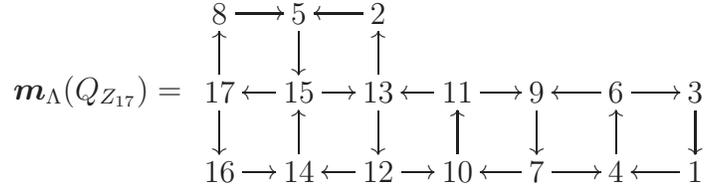
\begin{gathered}
\begin{xy} 0;<0.6pt,0pt>:<0pt,-0.6pt>:: 
(0,100) *+{16} ="0",
(0,0) *+{8} ="1",
(0,50) *+{17} ="2",
(50,100) *+{14} ="3",
(50,0) *+{5} ="4",
(50,50) *+{15} ="5",
(100,100) *+{12} ="6",
(100,0) *+{2} ="7",
(100,50) *+{13} ="8",
(150,100) *+{10} ="9",
(150,50) *+{11} ="10",
(200,100) *+{7} ="11",
(200,50) *+{9} ="12",
(250,100) *+{4} ="13",
(250,50) *+{6} ="14",
(300,100) *+{1} ="15",
(300,50) *+{3} ="16",
"2", {\ar"0"},
"0", {\ar"3"},
"2", {\ar"1"},
"1", {\ar"4"},
"5", {\ar"2"},
"3", {\ar"5"},
"6", {\ar"3"},
"4", {\ar"5"},
"7", {\ar"4"},
"5", {\ar"8"},
"8", {\ar"6"},
"6", {\ar"9"},
"8", {\ar"7"},
"10", {\ar"8"},
"9", {\ar"10"},
"11", {\ar"9"},
"10", {\ar"12"},
"12", {\ar"11"},
"11", {\ar"13"},
"14", {\ar"12"},
"13", {\ar"14"},
"15", {\ar"13"},
"14", {\ar"16"},
"16", {\ar"15"},
\end{xy}
\end{gathered}$$
\item type: $(A_7, c^4 : A_7, c^4  : A_3, c^2)$
\end{itemize}
\item[(b)] $(A_3 \times 3, A_2 \times 4)$ chamber:
\begin{itemize}
\item complete family: $A_3 = \{ 1, 3, 2 \} + 3k, k = 0,1,2$ $A_2 = \{10,11 \} + 2 h, h = 0,1,2,3$
\item Weyl--factorized sequence: 1 2 6 8 7 11 12 15 16 3 4 5 9 10 13 14 17 2 1 6 7 8 11 12 15 16 3 5 4 9 10 10
\item permutation:
$$ \boldsymbol{m}_{\Lambda} (Q_{ Z_{17 }})=
\begin{gathered}
\begin{xy} 0;<0.6pt,0pt>:<0pt,-0.6pt>:: 
(0,100) *+{2} ="0",
(0,0) *+{1} ="1",
(0,50) *+{3} ="2",
(50,100) *+{5} ="3",
(50,0) *+{4} ="4",
(50,50) *+{6} ="5",
(100,100) *+{8} ="6",
(100,0) *+{7} ="7",
(100,50) *+{9} ="8",
(150,100) *+{11} ="9",
(150,50) *+{10} ="10",
(200,100) *+{13} ="11",
(200,50) *+{12} ="12",
(250,100) *+{15} ="13",
(250,50) *+{14} ="14",
(300,100) *+{17} ="15",
(300,50) *+{16} ="16",
"2", {\ar"0"},
"0", {\ar"3"},
"2", {\ar"1"},
"1", {\ar"4"},
"5", {\ar"2"},
"3", {\ar"5"},
"6", {\ar"3"},
"4", {\ar"5"},
"7", {\ar"4"},
"5", {\ar"8"},
"8", {\ar"6"},
"6", {\ar"9"},
"8", {\ar"7"},
"10", {\ar"8"},
"9", {\ar"10"},
"11", {\ar"9"},
"10", {\ar"12"},
"12", {\ar"11"},
"11", {\ar"13"},
"14", {\ar"12"},
"13", {\ar"14"},
"15", {\ar"13"},
"14", {\ar"16"},
"16", {\ar"15"},
\end{xy}
\end{gathered}$$
\item type: $A_3, c^2$, $A_2, s_2 s_1 s_2$ for even k, $A_2, s_1 s_2 s_1$ for odd k.
\end{itemize}
\end{itemize}

\item ${\bf Z_{18}}$

\smallskip

In this case is more convenient to use another representative of $Z_{18}$ under $2d$ wall--crossing. Switching the vacuas $|17 \rangle$ and $|18 \rangle$, \emph{i.e.} using the $\alpha_{17}$ operation of \eqref{2dWX}, we bring the Coxeter--Dynkin graph of table \ref{ArnB1} in the form

{\tiny $$\begin{gathered}\xymatrix{
\bullet \ar@{-}[r]\ar@{-}[d]&\bullet \ar@{-}[r]\ar@{-}[d]\ar@{..}[dl]&\bullet \ar@{-}[d]\ar@{..}[dl] & &\\
\bullet \ar@{-}[r]\ar@{-}[d]\ar@{..}[dr]&\bullet \ar@{-}[r]\ar@{-}[d]\ar@{..}[dr]&\bullet \ar@{-}[r]\ar@{-}[d]\ar@{..}[dr]&\bullet \ar@{-}[r]\ar@{-}[d]\ar@{..}[dr]&\bullet \ar@{-}[r]\ar@{-}[d]\ar@{..}[dr]&\bullet \ar@{-}[d] \ar@{..}[dr]\ar@{-}[r]&18 \ar@{-}[d]\ar@{-}[dr]\\
\bullet \ar@{-}[r]&\bullet \ar@{-}[r]&\bullet \ar@{-}[r]&\bullet \ar@{-}[r]&\bullet \ar@{-}[r]&\bullet \ar@{-}[r]&\bullet&17\\
}\end{gathered}$$}

to this form we apply the method of section \S \ref{CDZMethod} to find a quiver representative of the $4d$ theory. The square form of this quiver is:
$$Q_{Z_{17}} \colon \begin{gathered}
\begin{xy} 0;<0.6pt,0pt>:<0pt,-0.6pt>:: 
(0,100) *+{1} ="0",
(0,0) *+{2} ="1",
(0,50) *+{3} ="2",
(50,100) *+{4} ="3",
(50,0) *+{5} ="4",
(50,50) *+{6} ="5",
(100,100) *+{7} ="6",
(100,0) *+{8} ="7",
(100,50) *+{9} ="8",
(150,100) *+{10} ="9",
(150,50) *+{11} ="10",
(200,100) *+{12} ="11",
(200,50) *+{13} ="12",
(250,100) *+{14} ="13",
(250,50) *+{15} ="14",
(300,100) *+{16} ="15",
(350,100) *+{17} ="16",
(300,50) *+{18} ="17",
"0", {\ar"2"},
"3", {\ar"0"},
"1", {\ar"2"},
"4", {\ar"1"},
"2", {\ar"5"},
"5", {\ar"3"},
"3", {\ar"6"},
"5", {\ar"4"},
"4", {\ar"7"},
"8", {\ar"5"},
"6", {\ar"8"},
"9", {\ar"6"},
"7", {\ar"8"},
"8", {\ar"10"},
"10", {\ar"9"},
"9", {\ar"11"},
"12", {\ar"10"},
"11", {\ar"12"},
"13", {\ar"11"},
"12", {\ar"14"},
"14", {\ar"13"},
"13", {\ar"15"},
"17", {\ar"14"},
"15", {\ar"16"},
"15", {\ar"17"},
\end{xy}
\end{gathered}
$$
To obtain the original form use the sequence: (17 18 15 13 11 16 14 18 11 2 1 5 4 3 7 6 8 2 1$)^{-1}$. Now, obtaining the chambers of equation \eqref{Z18ch} is straightforward: one proceeds as for $E_{19}$ using our results about $Z_{17}$. The more convenient Weyl--factorized sink--sequence to generate the $Z_{17}$ $Y$--system is the one associated to the chamber $(A_3 \times 3,A_2 \times 4, A_1)$. This is a chamber of $\mu_{17}(Q_{Z_{17}})$, with the obvious complete family of Dynkin's. The sequence is: 3 5 4 9 10 13 14 18 2 1 6 8 7 11 12 15 17 16 3 5 4 9 10 13 14 18 2 1 6 8 7.

\item ${\bf Z_{19}}$ and ${\bf Z_{1,0}}$

\smallskip

These two theories are analogous to $Z_{17}$. Again we list only the more convenient sink sequence for generating the $Y$--system.

\medskip

The square form representative for $Z_{19}$ is obtained with the sequence of mutations 2 1 5 4 3 19 17 18 15 16 19 13 14 17 18 11 12 15 16 19 from the one obtained with \S \ref{CDZMethod}. The result is:
$$ Q_{ Z_{19}} \colon \begin{gathered}
\begin{xy} 0;<0.6pt,0pt>:<0pt,-0.6pt>:: 
(0,100) *+{1} ="0",
(0,0) *+{2} ="1",
(0,50) *+{3} ="2",
(50,100) *+{4} ="3",
(50,0) *+{5} ="4",
(50,50) *+{6} ="5",
(100,100) *+{7} ="6",
(100,0) *+{8} ="7",
(100,50) *+{9} ="8",
(150,100) *+{10} ="9",
(150,50) *+{11} ="10",
(200,100) *+{12} ="11",
(200,50) *+{13} ="12",
(250,100) *+{14} ="13",
(250,50) *+{15} ="14",
(300,100) *+{16} ="15",
(300,50) *+{17} ="16",
(350,100) *+{18} ="17",
(350,50) *+{19} ="18",
"2", {\ar"0"},
"0", {\ar"3"},
"2", {\ar"1"},
"1", {\ar"4"},
"5", {\ar"2"},
"3", {\ar"5"},
"6", {\ar"3"},
"4", {\ar"5"},
"7", {\ar"4"},
"5", {\ar"8"},
"8", {\ar"6"},
"6", {\ar"9"},
"8", {\ar"7"},
"10", {\ar"8"},
"9", {\ar"10"},
"11", {\ar"9"},
"10", {\ar"12"},
"12", {\ar"11"},
"11", {\ar"13"},
"14", {\ar"12"},
"13", {\ar"14"},
"15", {\ar"13"},
"14", {\ar"16"},
"16", {\ar"15"},
"15", {\ar"17"},
"18", {\ar"16"},
"17", {\ar"18"},
\end{xy}
\end{gathered}$$
The sequence of the chamber $(A_3 \times 3 , A_2 \times 5)$ is: 2 1 6 8 7 11 12 15 16 19 3 5 4 9 10 13 14 17 18 2 1 6 7 8 11 12 15 16 19 3 5 4 9.

\medskip

The $Z_{1,0}$ case is analogous to this one. The square form of the quiver is obtained by removing from the previous one the nodes 16 17 18 19. The transformation that maps it back to the form we deduce from the Coxeter--Dynkin diagram is (2 1 5 4 3 15 13 14 11 12 15$)^{-1}$. The sequence of the chamber $(A_3 \times 3 , A_2 \times 3)$ is: 2 1 6 8 7 11 12 15 3 5 4 9 10 13 14 2 1 6 8 7 11 12 15 3 5 4 9.

\item ${\bf W_{17}}$

\smallskip

The study of the stable BPS hypermultiplets of this model is analogous to that of the odd rank elements of the $Z$ family. The square form of the quiver $Q_{W_{17}}$ is
$$ Q_{ W_{17}} \colon \begin{gathered}
\begin{xy} 0;<0.6pt,0pt>:<0pt,-0.6pt>:: 
(0,100) *+{1} ="0",
(0,0) *+{2} ="1",
(0,50) *+{3} ="2",
(50,100) *+{4} ="3",
(50,0) *+{5} ="4",
(50,50) *+{6} ="5",
(100,100) *+{7} ="6",
(100,0) *+{8} ="7",
(100,50) *+{9} ="8",
(150,100) *+{10} ="9",
(150,0) *+{11} ="10",
(150,50) *+{12} ="11",
(200,100) *+{13} ="12",
(200,0) *+{14} ="13",
(200,50) *+{15} ="14",
(250,100) *+{16} ="15",
(250,50) *+{17} ="16",
"0", {\ar"2"},
"3", {\ar"0"},
"1", {\ar"2"},
"4", {\ar"1"},
"2", {\ar"5"},
"5", {\ar"3"},
"3", {\ar"6"},
"5", {\ar"4"},
"4", {\ar"7"},
"8", {\ar"5"},
"6", {\ar"8"},
"9", {\ar"6"},
"7", {\ar"8"},
"10", {\ar"7"},
"8", {\ar"11"},
"11", {\ar"9"},
"9", {\ar"12"},
"11", {\ar"10"},
"10", {\ar"13"},
"14", {\ar"11"},
"12", {\ar"14"},
"15", {\ar"12"},
"13", {\ar"14"},
"14", {\ar"16"},
"16", {\ar"15"},
\end{xy}
\end{gathered}$$

It is related to the original form by the sequence of mutations: ( 1 2 5 4 8 7 11 10 13 3 6 9 12 2 1 4 5 8 7 14 3 6 2 1$)^{-1}$. The two chambers of \eqref{W17ch} are straightforwardly generated. The more convenient to generate the $Y$--system is the $(A_3 \times 5, A_2)$ one for which the Weyl--factorized sequence is: 
3 5 4 9 11 10 15 16 2 1 6 8 7 12 14 13 17 3 5 4 9 11 10 15 2 1 6 8 7 12 13 14 16.

\item ${\bf Q_{2,0}}, {\bf Q_{16}} \text{ and } {\bf Q_{18}}$

\smallskip

We will discuss $Q_{2,0}$ as an example for all the theories in the $Q$ family of even rank. The square form of $Q_{2,0}$ is
$$ Q_{ Q_{2,0 }} \colon 
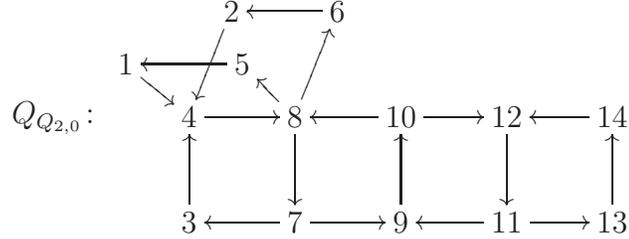
\begin{gathered}
\begin{xy} 0;<0.8pt,0pt>:<0pt,-0.8pt>:: 
(0,25) *+{1} ="0",
(50,0) *+{2} ="1",
(30,100) *+{3} ="2",
(30,50) *+{4} ="3",
(55,25) *+{5} ="4",
(100,0) *+{6} ="5",
(80,100) *+{7} ="6",
(80,50) *+{8} ="7",
(130,100) *+{9} ="8",
(130,50) *+{10} ="9",
(180,100) *+{11} ="10",
(180,50) *+{12} ="11",
(230,100) *+{13} ="12",
(230,50) *+{14} ="13",
"0", {\ar"3"},
"4", {\ar"0"},
"1", {\ar"3"},
"5", {\ar"1"},
"2", {\ar"3"},
"6", {\ar"2"},
"3", {\ar"7"},
"7", {\ar"4"},
"7", {\ar"5"},
"7", {\ar"6"},
"6", {\ar"8"},
"9", {\ar"7"},
"8", {\ar"9"},
"10", {\ar"8"},
"9", {\ar"11"},
"11", {\ar"10"},
"10", {\ar"12"},
"13", {\ar"11"},
"12", {\ar"13"},
\end{xy}
\end{gathered}$$

To bring it back to its original form: (1 2 3 14 12 10 13 11 14$)^{-1}$. The model has two canonical BPS chambers:
\begin{itemize}
\item[(a)] $(D_4 \times 2, A_2 \times 3)$ chamber:
\begin{itemize}
\item complete family: $D_4 = \{ 1 , 2, 3, 4 \} + 4 k, k = 0, 1$ $A_2 = \{ 9, 10 \} + 2 n, n = 0,1,2$.
\item Weyl--factorized sequence: 4 5 6 7 10 11 14 1 2 3 8 9 12 13 4 5 6 7 1 2 3 8 4 5 6 7 10 11 14 1 2 3 8
\item permutation: 
$$ \boldsymbol{m}_{\Lambda} (Q_{ Q_{2,0 }})=
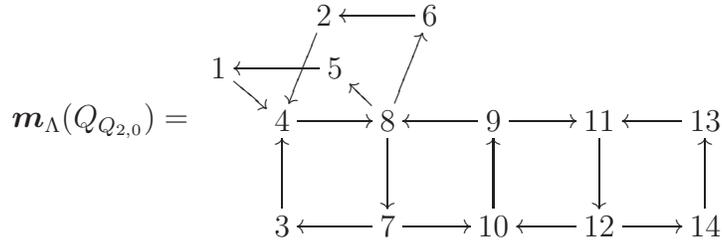
\begin{gathered}
 \begin{xy} 0;<0.8pt,0pt>:<0pt,-0.8pt>:: 
(0,25) *+{1} ="0",
(50,0) *+{2} ="1",
(30,100) *+{3} ="2",
(30,50) *+{4} ="3",
(55,25) *+{5} ="4",
(100,0) *+{6} ="5",
(80,100) *+{7} ="6",
(80,50) *+{8} ="7",
(130,100) *+{10} ="8",
(130,50) *+{9} ="9",
(180,100) *+{12} ="10",
(180,50) *+{11} ="11",
(230,100) *+{14} ="12",
(230,50) *+{13} ="13",
"0", {\ar"3"},
"4", {\ar"0"},
"1", {\ar"3"},
"5", {\ar"1"},
"2", {\ar"3"},
"6", {\ar"2"},
"3", {\ar"7"},
"7", {\ar"4"},
"7", {\ar"5"},
"7", {\ar"6"},
"6", {\ar"8"},
"9", {\ar"7"},
"8", {\ar"9"},
"10", {\ar"8"},
"9", {\ar"11"},
"11", {\ar"10"},
"10", {\ar"12"},
"13", {\ar"11"},
"12", {\ar"13"},
\end{xy}
\end{gathered}$$
\item type: $D_4, c^3 : D_4, c^3$, while $A_2, s_2 s_1 s_2$ for $n= 0, 2$, $A_2, s_1 s_2 s_1$ for $n=1$.
\end{itemize}
\item[(b)] $(A_2 \times 2, A_5 \times 2)$ chamber:
\begin{itemize}
\item complete family: $A_2 = \{1,5 \}, \{2 , 6 \}$ $A_5 = \{ 3,7,10,11,13\} + n, n= 0,1$
\item Weyl--factorized sequence: 2 1 8 12 3 10 13 4 9 14 7 11 8 12 3 10 13 4 9 14 7 11 8 12 3 10 13 5 6 4 9 14 7 11 1 2
\item permutation: 
$$ \boldsymbol{m}_{\Lambda} (Q_{ Q_{2,0 }})=
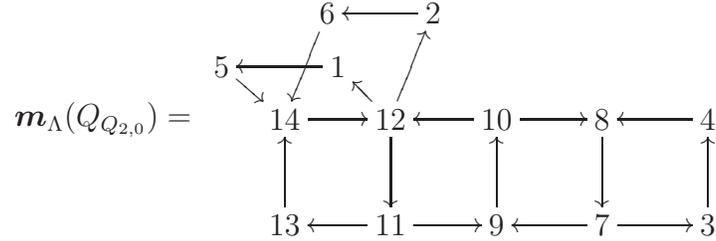
\begin{gathered}
\begin{xy} 0;<0.8pt,0pt>:<0pt,-0.8pt>:: 
(0,25) *+{5} ="0",
(50,0) *+{6} ="1",
(30,100) *+{13} ="2",
(30,50) *+{14} ="3",
(55,25) *+{1} ="4",
(100,0) *+{2} ="5",
(80,100) *+{11} ="6",
(80,50) *+{12} ="7",
(130,100) *+{9} ="8",
(130,50) *+{10} ="9",
(180,100) *+{7} ="10",
(180,50) *+{8} ="11",
(230,100) *+{3} ="12",
(230,50) *+{4} ="13",
"0", {\ar"3"},
"4", {\ar"0"},
"1", {\ar"3"},
"5", {\ar"1"},
"2", {\ar"3"},
"6", {\ar"2"},
"3", {\ar"7"},
"7", {\ar"4"},
"7", {\ar"5"},
"7", {\ar"6"},
"6", {\ar"8"},
"9", {\ar"7"},
"8", {\ar"9"},
"10", {\ar"8"},
"9", {\ar"11"},
"11", {\ar"10"},
"10", {\ar"12"},
"13", {\ar"11"},
"12", {\ar"13"},
\end{xy}
\end{gathered}$$
\item type: $A_5, c^3$, $A_2, s_1 s_2 s_1$.
\end{itemize}
\end{itemize}

The situation we encountered in chamber (a) is typical for all the even $Q$'s: the $D_4$ part remains invariant while the $A_2$ part gets switched. Proceeding analogously one obtains the chambers for $Q_{16}$ and $Q_{18}$ we listed in \eqref{Q16ch} and \eqref{Q18ch}.

\item ${\bf Q_{17}}$

For the $Q_{17}$ model we have to use the $2d$ wall--crossing group to generate the following configurations of $2d$ vacua and interpolating BPS states:

{\tiny 
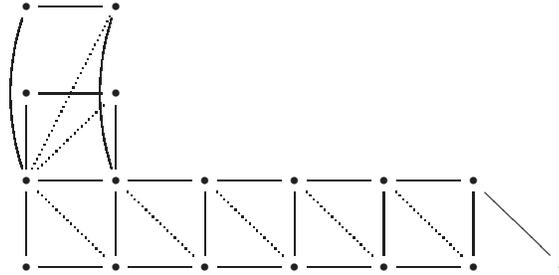
$$\begin{gathered}\xymatrix{\bullet\ar@{-}[r]\ar@{-}@/_/[dd]&\bullet \ar@{..}[ddl]\ar@{-}@/_/[dd]&&\\
\bullet \ar@{-}[r]\ar@{-}[d]&\bullet \ar@{-}[d]\ar@{..}[dl]& &\\
\bullet \ar@{-}[r]\ar@{-}[d]\ar@{..}[dr]&\bullet \ar@{-}[r]\ar@{-}[d]\ar@{..}[dr]&\bullet \ar@{-}[r]\ar@{-}[d]\ar@{..}[dr]&\bullet \ar@{-}[r]\ar@{-}[d]\ar@{..}[dr]&\bullet \ar@{-}[r]\ar@{-}[d]\ar@{..}[dr]&\bullet\ar@{-}[d]\ar@{-}[dr] \\
\bullet \ar@{-}[r]&\bullet \ar@{-}[r]&\bullet \ar@{-}[r]&\bullet \ar@{-}[r]&\bullet \ar@{-}[r]&\bullet&\bullet\\
}\end{gathered}$$ }

From this configuration we obtain a quiver with potential that by the sequence of mutations 1 2 3 7 4 5 6  17 16 14 12 15 13 16 reduces to the square form:
$$ Q_{ Q_{17 }} \colon 
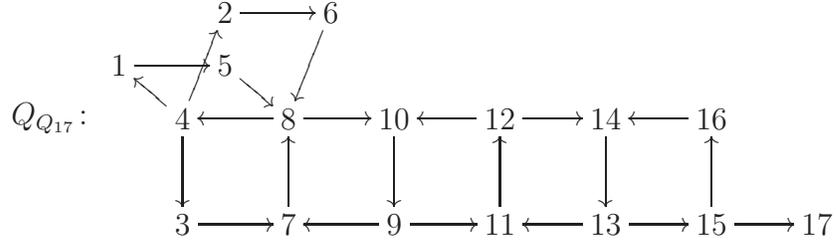
\begin{gathered}
\begin{xy} 0;<0.8pt,0pt>:<0pt,-0.8pt>:: 
(0,25) *+{1} ="0",
(50,0) *+{2} ="1",
(30,100) *+{3} ="2",
(30,50) *+{4} ="3",
(50,25) *+{5} ="4",
(100,0) *+{6} ="5",
(80,100) *+{7} ="6",
(80,50) *+{8} ="7",
(130,100) *+{9} ="8",
(130,50) *+{10} ="9",
(180,100) *+{11} ="10",
(180,50) *+{12} ="11",
(230,100) *+{13} ="12",
(230,50) *+{14} ="13",
(280,100) *+{15} ="14",
(280,50) *+{16} ="15",
(330,100) *+{17} ="16",
"3", {\ar"0"},
"0", {\ar"4"},
"3", {\ar"1"},
"1", {\ar"5"},
"3", {\ar"2"},
"2", {\ar"6"},
"7", {\ar"3"},
"4", {\ar"7"},
"5", {\ar"7"},
"6", {\ar"7"},
"8", {\ar"6"},
"7", {\ar"9"},
"9", {\ar"8"},
"8", {\ar"10"},
"11", {\ar"9"},
"10", {\ar"11"},
"12", {\ar"10"},
"11", {\ar"13"},
"13", {\ar"12"},
"12", {\ar"14"},
"15", {\ar"13"},
"14", {\ar"15"},
"14", {\ar"16"},
\end{xy}
\end{gathered}$$

This quiver is a one--point extension of $Q_{16}$. From this fact our results about the BPS chambers follows as in the $E_{19}$ example. The shortest sequence to generate the $Y$--system is the one associated to the $(D_4\times 2, A_2 \times 4 ,A_1)$ chamber of $\mu_{17} (Q_{17})$: 1 2 3 8 9 12 13 16 4 5 6 7 1 2 3 8 4 5 6 7 1 2 3 8 10 11 14 17 15 4 5 6 7 9 12 13 16.

\item ${\bf S_{17}}$, $\bf S_{16}$ and ${\bf S_{1,0}}$

The square form of the quiver $S_{17}$ is:
$$ Q_{S_{17 }} \colon 
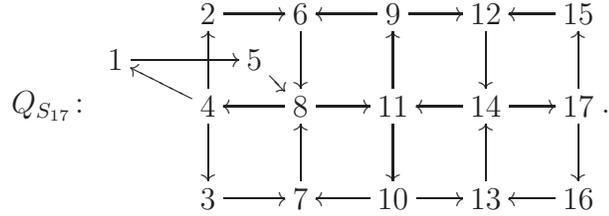
\begin{gathered}
\begin{xy} 0;<0.7pt,0pt>:<0pt,-0.7pt>:: 
(0,25) *+{1} ="0",
(50,0) *+{2} ="1",
(50,100) *+{3} ="2",
(50,50) *+{4} ="3",
(75,25) *+{5} ="4",
(100,0) *+{6} ="5",
(100,100) *+{7} ="6",
(100,50) *+{8} ="7",
(150,0) *+{9} ="8",
(150,100) *+{10} ="9",
(150,50) *+{11} ="10",
(200,0) *+{12} ="11",
(200,100) *+{13} ="12",
(200,50) *+{14} ="13",
(250,0) *+{15} ="14",
(250,100) *+{16} ="15",
(250,50) *+{17} ="16",
"3", {\ar"0"},
"0", {\ar"4"},
"3", {\ar"1"},
"1", {\ar"5"},
"3", {\ar"2"},
"2", {\ar"6"},
"7", {\ar"3"},
"4", {\ar"7"},
"5", {\ar"7"},
"8", {\ar"5"},
"6", {\ar"7"},
"9", {\ar"6"},
"7", {\ar"10"},
"10", {\ar"8"},
"8", {\ar"11"},
"10", {\ar"9"},
"9", {\ar"12"},
"13", {\ar"10"},
"11", {\ar"13"},
"14", {\ar"11"},
"12", {\ar"13"},
"15", {\ar"12"},
"13", {\ar"16"},
"16", {\ar"14"},
"16", {\ar"15"},
\end{xy}
\end{gathered}.$$
It is related with the Coxeter--Dynkin graph by the sequence of mutations ( 17 14 15 16 2 1 3 6 7 4 5 $)^{-1}$. The two chambers we listed in \eqref{S17ch} are easily obtained with the help of the Keller's applet.
\begin{itemize}
\item[(a)] $(D_4 \times 2, A_3 \times 3)$ chamber
\begin{itemize}
\item complete family: $D_4 = \{1,2,3,4\} + 4k, k=0,1$ $A_3 = \{ 9,10,11\} + 3m, m = 0,1,2$
\item Weyl--factorized sequence: 1 2 3 8 9 10 14 15 16 4 5 6 7 11 12 13 17 1 2 3 8 4 5 6 7 9 10 14 15 16 1 2 3 8 4 5 6 7 11 12 13 17
\item permutation:
$$ \boldsymbol{m}_{\Lambda} (Q_{ S_{17}})=
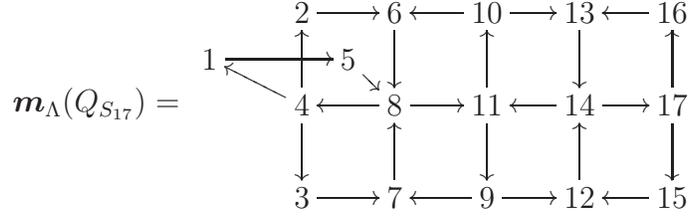
\begin{gathered}
\begin{xy} 0;<0.7pt,0pt>:<0pt,-0.7pt>:: 
(0,25) *+{1} ="0",
(50,0) *+{2} ="1",
(50,100) *+{3} ="2",
(50,50) *+{4} ="3",
(75,25) *+{5} ="4",
(100,0) *+{6} ="5",
(100,100) *+{7} ="6",
(100,50) *+{8} ="7",
(150,0) *+{10} ="8",
(150,100) *+{9} ="9",
(150,50) *+{11} ="10",
(200,0) *+{13} ="11",
(200,100) *+{12} ="12",
(200,50) *+{14} ="13",
(250,0) *+{16} ="14",
(250,100) *+{15} ="15",
(250,50) *+{17} ="16",
"3", {\ar"0"},
"0", {\ar"4"},
"3", {\ar"1"},
"1", {\ar"5"},
"3", {\ar"2"},
"2", {\ar"6"},
"7", {\ar"3"},
"4", {\ar"7"},
"5", {\ar"7"},
"8", {\ar"5"},
"6", {\ar"7"},
"9", {\ar"6"},
"7", {\ar"10"},
"10", {\ar"8"},
"8", {\ar"11"},
"10", {\ar"9"},
"9", {\ar"12"},
"13", {\ar"10"},
"11", {\ar"13"},
"14", {\ar"11"},
"12", {\ar"13"},
"15", {\ar"12"},
"13", {\ar"16"},
"16", {\ar"14"},
"16", {\ar"15"},
\end{xy}
\end{gathered}$$
\item type: $D_4,c^3$ and $A_3,c^2$
\end{itemize}
\item[(b)] $(A_5 \times 3 , A_2)$ chamber
\begin{itemize}
\item complete family: $A_5 = \{ 2,6,10,13,16 \}, \{4,8,11,14,17\}, \{3,7,9,12,15\}$ $A_2 = \{1,5\}$
\item Weyl--factorized sequence: 5 6 12 4 11 17 7 13 1 2 9 15 8 14 3 10 16 6 12 4 11 17 7 13 2 9 15 8 14 3 10 16 6 12 4 11 17 7 13 2 9 15 8 14 3 10 16 5
\item permutation:
$$ \boldsymbol{m}_{\Lambda} (Q_{ S_{17}})=
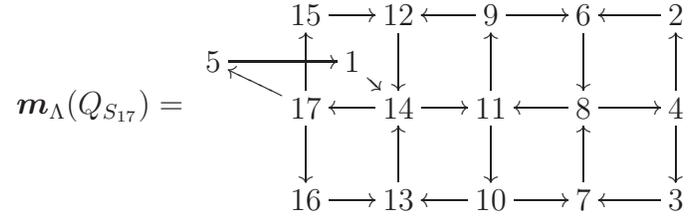
\begin{gathered}
\begin{xy} 0;<0.7pt,0pt>:<0pt,-0.7pt>:: 
(0,25) *+{5} ="0",
(50,0) *+{15} ="1",
(50,100) *+{16} ="2",
(50,50) *+{17} ="3",
(75,25) *+{1} ="4",
(100,0) *+{12} ="5",
(100,100) *+{13} ="6",
(100,50) *+{14} ="7",
(150,0) *+{9} ="8",
(150,100) *+{10} ="9",
(150,50) *+{11} ="10",
(200,0) *+{6} ="11",
(200,100) *+{7} ="12",
(200,50) *+{8} ="13",
(250,0) *+{2} ="14",
(250,100) *+{3} ="15",
(250,50) *+{4} ="16",
"3", {\ar"0"},
"0", {\ar"4"},
"3", {\ar"1"},
"1", {\ar"5"},
"3", {\ar"2"},
"2", {\ar"6"},
"7", {\ar"3"},
"4", {\ar"7"},
"5", {\ar"7"},
"8", {\ar"5"},
"6", {\ar"7"},
"9", {\ar"6"},
"7", {\ar"10"},
"10", {\ar"8"},
"8", {\ar"11"},
"10", {\ar"9"},
"9", {\ar"12"},
"13", {\ar"10"},
"11", {\ar"13"},
"14", {\ar"11"},
"12", {\ar"13"},
"15", {\ar"12"},
"13", {\ar"16"},
"16", {\ar"14"},
"16", {\ar"15"},
\end{xy}
\end{gathered}$$
\item type $A_5:c^3$, $A_2: s_2 s_1 s_2$
\end{itemize}
\end{itemize}
The cases of $S_{16}$ and $S_{1,0}$ can be treated analogously. The square form of the quiver $S_{16}$ is the same as $Q_{S_{17}}$ with the node 15 removed, and consequent relabeling (17 $\rightarrow$ 16 and 16 $\rightarrow$ 15), while the quiver of $S_{1,0}$ is obtained by the removal of the nodes 15 16 17. As far as $S_{1,0}$ the shortest sequence to obtain the $Y$--system is the one leading to the $(D_4 \times 2, A_3 \times 2)$ chamber: 1 2 3 8 9 10 14 4 5 6 7 11 12 13 1 2 3 8 4 5 6 7 1 2 3 8 9 10 14 4 6 7 11 12 13 5. For $S_{16}$ it is the one leading to the $(D_4 \times 2, A_3\times 2, A_2)$ chamber: 1 2 3 8 9 10 14 15 4 5 6 7 11 12 13 16 1 2 3 8 4 5 6 7 1 2 3 8 9 10 14 4 5 7 6 11 12 13 15.

\item ${\bf U_{1,0}}$
The square form quiver representative of the $U_{1,0}$ theory is the following:
$$ Q_{ U_{1,0 }} \colon 
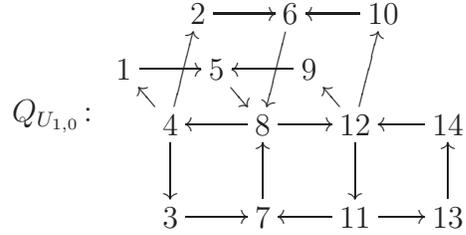
\begin{gathered}
\begin{xy} 0;<0.7pt,0pt>:<0pt,-0.7pt>:: 
(0,30) *+{1} ="0",
(40,0) *+{2} ="1",
(25,110) *+{3} ="2",
(25,60) *+{4} ="3",
(50,30) *+{5} ="4",
(90,0) *+{6} ="5",
(75,110) *+{7} ="6",
(75,60) *+{8} ="7",
(100,30) *+{9} ="8",
(140,0) *+{10} ="9",
(125,110) *+{11} ="10",
(125,60) *+{12} ="11",
(175,110) *+{13} ="12",
(175,60) *+{14} ="13",
"3", {\ar"0"},
"0", {\ar"4"},
"3", {\ar"1"},
"1", {\ar"5"},
"3", {\ar"2"},
"2", {\ar"6"},
"7", {\ar"3"},
"4", {\ar"7"},
"8", {\ar"4"},
"5", {\ar"7"},
"9", {\ar"5"},
"6", {\ar"7"},
"10", {\ar"6"},
"7", {\ar"11"},
"11", {\ar"8"},
"11", {\ar"9"},
"11", {\ar"10"},
"10", {\ar"12"},
"13", {\ar"11"},
"12", {\ar"13"},
\end{xy}
\end{gathered}$$
Related to the one obtained with the method of \S 4.1 by the sequence 14 4 9 10 11 13. The analysis of the algebraically trivial chambers of this theory is straightforward. The shortest sink--sequence is the one associated to the $(D_4 \times 3, A_2)$ chamber that reads: 1 2 3 8 9 10 11 14 4 5 6 7 12 13 1 2 3 8 9 10 11 4 5 6 7 12 2 1 3 8 9 10 11 14 4 5 6 7 12.
\end{itemize}

From the above examples, obtaining the Weyl--factorized sequences for the other chambers we listed in \S \ref{BPSspectra} should now be an easy exercise with the help of Keller's applet \cite{applet}.

\newpage


\end{document}